\documentclass[letterpaper,twocolumn,10pt]{article}
\usepackage{usenix-2020-09}
\usepackage{cite}
\usepackage{amsmath,amssymb,amsfonts}
\usepackage{algorithmic}
\usepackage{algorithm}
\usepackage{graphicx}
\usepackage{textcomp}
\usepackage{xcolor}
\usepackage{xurl}
\usepackage{hyperref}
\usepackage{booktabs}
\usepackage{pifont}
\usepackage{adjustbox}
\usepackage{array}
\usepackage{multirow}
\usepackage{cancel}
\def\BibTeX{{\rm B\kern-.05em{\sc i\kern-.025em b}\kern-.08em
    T\kern-.1667em\lower.7ex\hbox{E}\kern-.125emX}}
\usepackage{comment}
\usepackage{todonotes}

\usepackage{soul}
\newlength\MAX  

\usepackage{pifont}% http://ctan.org/pkg/pifont
\newcommand{\cmark}{\ding{51}}%
\newcommand{\xmark}{\ding{55}}%

\usepackage[most]{tcolorbox}

\tcbset{
enhanced,
breakable,
left=2mm,
right=2mm,
attach boxed title to top left={
xshift=0.2cm,
yshift=-5mm,
yshifttext=-1mm
},
top=4mm,
colback=blue!5!white,
colframe=blue!100!black,
colbacktitle=blue!100!black,
boxed title style={size=small,colframe=blue!100!black}
}
\newcommand{\DrawPercentageBar}[1]{%
  \begin{tikzpicture}
    \fill[color=black]   (0.0 , 0.0) rectangle (#1*3ex , 1.5ex );
    \fill[color=gray] (#1*3ex  , 0.0) rectangle (3.0ex, 1.5ex);
  \end{tikzpicture}%
}

\newcommand{\DrawPercentageBarBlue}[1]{%
  \begin{tikzpicture}
    \fill[color=blue]   (0.0 , 0.0) rectangle (#1*3ex , 1.5ex );
    \fill[color=gray] (#1*3ex  , 0.0) rectangle (3.0ex, 1.5ex);
  \end{tikzpicture}%
}
\usepackage{setspace}
\usepackage{hyperref}
\newenvironment{mybox}[1]{%
\begin{tcolorbox}[title={#1}]%
\setstretch{0.95}}{
\end{tcolorbox}
}

\usepackage{tikz}
\usepackage{amsmath}

\usepackage{filecontents}

\begin{document}
%-------------------------------------------------------------------------------

%don't want date printed
\date{}

% make title bold and 14 pt font (Latex default is non-bold, 16 pt)
\title{\Large \bf The Power of Words: Generating PowerShell Attacks from Natural Language}

%\author{\textit{Anonymous authors for double-blind review}}

%\begin{comment}

%for single author (just remove % characters)
\author{
{\rm Pietro Liguori*, Christian Marescalco**, Roberto Natella*, Vittorio Orbinato*, Luciano Pianese*}\\
DIETI, Università degli Studi di Napoli Federico II, Naples, Italy\\
*\{pietro.liguori, roberto.natella, vittorio.orbinato, luciano.pianese\}@unina.it\\
**c.marescalco@studenti.unina.it
} % end author

\begin{comment}
\and
{\rm Christian Marescalco}\\
DIETI, Università degli Studi di Napoli Federico II, Naples, Italy
\and
{\rm Roberto Natella}\\
DIETI, Università degli Studi di Napoli Federico II, Naples, Italy
\and
{\rm Vittorio Orbinato}\\
DIETI, Università degli Studi di Napoli Federico II, Naples, Italy
\and
{\rm Luciano Pianese}\\
DIETI, Università degli Studi di Napoli Federico II, Naples, Italy
% copy the following lines to add more authors
% \and
% {\rm Name}\\
%Name Institution
%} % end author

\end{comment}

\maketitle

\begin{abstract}
As the Windows OS stands out as one of the most targeted systems, the \textit{PowerShell} language has become a key tool for malicious actors and cybersecurity professionals (e.g., for penetration testing).  
This work explores an uncharted domain in AI code generation by automatically generating offensive PowerShell code from natural language descriptions using Neural Machine Translation (NMT). 
For training and evaluation purposes, we propose two novel datasets with PowerShell code samples, one with manually curated descriptions in natural language and another code-only dataset for reinforcing the training. We present an extensive evaluation of state-of-the-art NMT models and analyze the generated code both statically and dynamically. 
Results indicate that tuning NMT using our dataset is effective at generating offensive PowerShell code. Comparative analysis against the most widely used LLM service ChatGPT reveals the specialized strengths of our fine-tuned models. %for offensive PowerShell code generation. This work pioneers the automatic generation of offensive PowerShell code from natural language, offering practical insights and contributing valuable datasets to the research community.
\end{abstract}

\section{Introduction}
\label{sec:introduction}

\textit{Offensive security} practices, such as red teaming and adversary emulation, play a crucial role by helping us to understand how attackers take advantage of vulnerabilities and how to mitigate attacks~\cite{Applebaum2016IntelligentAR, ajmal2021offensive}.
In these attacks, cybersecurity professionals emulate malicious post-exploitation actions, such as credential stealing, lateral movement across accounts and machines, data obfuscation and exfiltration, and more~\cite{hutchins2011intelligence}. 

As Windows stands out as one of the most targeted OS~\cite{strom2018mitre}, the \textit{PowerShell} language has become a key tool for both malicious actors and cybersecurity professionals. This language is widely used to perform attacks since it can perform complex actions, such as establishing connections and accessing OS services and APIs without the need to deliver a malicious binary executable or payload on the target machine (e.g., ``fileless'' malware), making them harder to detect~\cite{sudhakar2020emerging, kara2023fileless, FilelessMalwarePowerShell,cisa}. 

Unfortunately, writing offensive code demands a high degree of expertise and effort, restricting the adoption of offensive security practices. Therefore, the rise of automatic \emph{AI code generators} represents an appealing solution to unlock these practices to a broader spectrum of users~\cite{natella2024ai}. 

AI code generators leverage ML models for Neural Machine Translation (NMT) to produce (offensive) code starting from inputs in Natural Language (NL), e.g., in the English language. 
The usage of NMT models is widespread across diverse software engineering tasks~\cite{fan2023large}, yet their application in security-related scenarios is infrequent and not widely explored. This gap stems primarily from the lack of suitable corpora for training and evaluating code generators. 
The shortage of corpora for offensive code generation is an evident limitation: existing benchmarks \cite{chen2021evaluating,yu2024codereval,du2023classeval} are derived from programming competitions and software interview questions (e.g., about algorithms and mathematics), or they focus on programs and languages that are not related to security (e.g., web applications in Python). 
Only a few security-oriented datasets are publicly available, targeting shellcodes in low-level programming languages~\cite{liguori-etal-2021-shellcode}. 
As a result, there is a significant gap in the literature on offensive PowerShell code generation. 
 
This work presents an assessment of AI code generators for PowerShell offensive code, a novel application of NMT. 
Given that generative models are predominantly trained on mainstream programming languages like Python and Java, we investigate strategies to repurpose these models for the PowerShell domain.
To this aim, we adopt a combination of unlabeled and labeled datasets to train and evaluate models. Specifically, we first use a large collection of unlabeled (i.e., code only) samples of general-purpose PowerShell from various online repositories to pre-train ML models and refine their capabilities to comprehend and generate PowerShell code. 
Then, we build from scratch a manually annotated labeled dataset consisting of PowerShell code samples specifically crafted for security applications, which we pair with curated NL descriptions in English. We use this dataset to fine-tune three state-of-the-art NMT models (CodeT5+ \cite{wang2023codet5+}, CodeGPT \cite{lu2021codexglue}, and CodeGen \cite{codegen}) to generate offensive PowerShell code. The dataset also serves as a ground truth for the evaluation. 
We publicly share code, models \footnote{\href{https://huggingface.co/collections/dessertlab/the-power-of-words-generating-powershell-attacks-from-natur-66223c3e6cd34bb31ce38a69}{HuggingFace repo}} and datasets as open data\footnote{\href{https://github.com/dessertlab/powershell-offensive-code-generation/}{GitHub repo}} to encourage further experimentation on this topic.

To perform our experiments, we formulate four key research questions (RQs) aimed at evaluating the models' capabilities and the impact of the training strategies, performing static and execution analysis to assess the generated code, 
and comparing privately fine-tuned models with ChatGPT, the most widely used LLM service from OpenAI~\cite{ChatGPT}. \tablename~\ref{tab:findings} summarizes the key findings of our analysis. 
To the best of our knowledge, this is the first work on the automatic generation of offensive PowerShell code from NL descriptions.

In the following, 
Section~\ref{sec:related} discusses related work;
Section~\ref{sec:research} describes the research study;
Section~\ref{sec:evaluation} shows the experimental results;
Section~\ref{sec:threats} discusses the threats to validity;
Section~\ref{sec:ethics} discusses the ethical considerations;
Section~\ref{sec:conclusion} concludes the paper.

\begin{table}[!ht]
\centering
\small
\begin{tabular}{
>{\centering\arraybackslash}m{1.5cm} |
>{\centering\arraybackslash}m{6.5cm}} 
\toprule
\textbf{Analysis} & \textbf{Main Findings} \\ \midrule
\textit{Capability Assessment} &
\begin{itemize}
  \item Models without fine-tuning (\textit{zero-shot learning}) showed a limited ability to generate PowerShell code, often defaulting to Python syntax or incorrect PowerShell code.
  \item The fine-tuning phase significantly enhanced the models' ability to generate syntactically correct and semantically relevant PowerShell code. Among the models, CodeT5+ and CodeGPT demonstrated notable improvements in generating offensive PowerShell code.
  \item Pre-training on a large PowerShell corpus had a varying impact on different models. While pre-training generally improved CodeT5+ and CodeGPT, especially with a limited number of epochs for fine-tuning, CodeGen did not consistently benefit from pre-training.
\end{itemize} \\ \midrule
\textit{Static and Execution Analysis} & \begin{itemize}
    \item All models achieved high syntax accuracy, indicating their strong capability to generate syntactically correct code. However, a significant number of warnings were identified, suggesting potential issues or suboptimal coding practices.
    \item The execution analysis showed that, despite textual differences between the ground truth and the generated code, the models are still able to generate offensive PowerShell code closely aligned with the intended malicious activities, in terms of events occurring in the system (e.g., on the filesystem, network, registry).
\end{itemize} \\
 \midrule
\textit{Comparison with public AI model} &
\begin{itemize}
  \item Our fine-tuned models outperform ChatGPT across all the metrics, showing that specializing the models on our fine-tuning dataset provides an advantage in the offensive PowerShell code generation task.
\end{itemize} \\
 \bottomrule
\end{tabular}
\caption{Main findings.}
\label{tab:findings}
\end{table}
\section{Related Work}
\label{sec:related}
This work focuses on offensive code generation, involving machine translation techniques applied to the security domain for PowerShell code generation. Thus, we reviewed related literature in these areas.

\vspace{3pt}
\noindent
\textbf{ML for security-related PowerShell.} Li \emph{et al.}~\cite{li2019effective} designed a subtree-based de-obfuscation method and a semantic-aware PowerShell attack detection system. This work also demonstrates how the presented de-obfuscation method improves the performance of detection systems such as Windows Defender and Virus-Total. PowerDP~\cite{PowerDP-2023} is a solution that aims to automatically identify malicious PowerShell commands through character distribution features and obfuscation multi-label classification also proposing a de-obfuscator method for recovering obfuscated commands.
Even ML-based methodologies have arisen for detection purposes, as shown by Hendler \emph{et al.} \cite{hendler2018detecting}, who proposed several ML-based detectors demonstrating their effectiveness on malicious scripts. The authors also devised another solution~\cite{rubin2019amsi} to achieve the same objective by retrieving information from Microsoft's AMSI interface. Mimura and Tajiri~\cite{MIMURA2021100404} presented a lighter methodology, restricting detection only to word embeddings. Mezawa \emph{et al.}~\cite{mezawa2022evaluating} proposed an evaluation methodology for ML-based detectors based on a word-level machine learning model. Given the effectiveness of Abstract Syntax Trees (ASTs) in detecting obfuscated PowerShell scripts, Rusak \emph{et al.}~\cite{rusak2018ast} proposed a hybrid approach that combines ASTs and deep learning to enhance detection methods for high-level obfuscation PowerShell malicious programs. 
We remark that research of ML for PowerShell focuses on \emph{defensive} uses (i.e., detecting and de-obfuscating attacks), but none of these studies analyzed the \emph{offensive} uses of ML (i.e., generating attacks), which are also relevant for red teaming and adversary emulation purposes, and which are in the scope of this paper.

\vspace{3pt}
\noindent
\textbf{Offensive Code Generation.} Research on AI code generators for offensive security is still at an early stage. Gupta \emph{et al.}~\cite{ThreatGPT} presented an outlook of the possibilities opened by ChatGPT for generating various types of cyber attacks, such as social engineering, phishing attacks, and malware creation. For each attack scenario, the paper shows qualitative examples of prompts submitted to ChatGPT, and the attack payloads generated as a result, including some snippets of PowerShell code. 
%investigated the possibilities provided by ChatGPT for composing various types of cyber attacks generating codes in many languages, ranging from Python to C, including some PowerShell snippets. Their research focuses on generating code for multiple scenarios, i.e., attack payloads, ransomware composition, and the generation of specific malware code. 
Similarly, Charan \emph{et al.}~\cite{charan2023text} presented qualitative examples with ChatGPT and Google BARD to generate malicious scripts (mainly in Python, Bash, and PowerShell) for the top 10 prevalent MITRE Techniques of 2022, showing the potential of these AI models for security applications. 
However, none of these studies systematically analyzed AI code generators, lacking in several aspects: (i) the evaluation was limited to a few examples, while systematic evaluation requires much larger datasets; (ii) the study lacked a ground truth for evaluating the correctness of generated code; (iii) they did not yet explore the potential of fine-tuning ML models for security-related code generation. 
The few studies in this direction focused on generating \emph{exploits} in low-level languages (e.g., to attack memory management vulnerabilities). However, exploitation is only a limited part of the cyber kill chain, overlooking several more types of malicious code. Among these studies, Liguori \emph{et al.}~\cite{liguori2022can} proposed a dataset and approach for training and evaluating AI code generators for code security, by generating shellcodes in Assembly language. EVIL~\cite{liguori2021evil} automatically generates exploits for conducting code injection attacks via NMT by targeting both the generation of shellcodes in Assembly language and related Python code for encoding and obfuscating the shellcodes. 
DualSC~\cite{yang2022dualsc} formalizes the automatic generation and summarization of shellcodes via a "Shallow" Transformer inspired by the T5 model and dual learning using the corpus provided by Liguori \emph{et al.}~\cite{liguori2022can}. 
ExploitGen~\cite{yang2023exploitgen} is an approach for generating exploit code in Python and Assembly based on the CodeBERT model. 
Differently from these studies, we presented a dedicated model for generating offensive PowerShell code, covering the entire cyber kill chain (e.g., including credential stealing, lateral movement, data exfiltration, and more tactics from the MITRE ATT\&CK taxonomy). Moreover, we systematically analyzed the quality of generated PowerShell code by introducing a manually curated dataset to serve as a ground truth and evaluating the code statically and dynamically.
\section{Research Study}
\label{sec:research}
%\roberto{NOTA PER LA FIGURA: dividiamo in tre parti il box di automatic evaluation, con similarity analysis, static analysis, dynamic analysis}

% Partiamo dall abilita di LLM di generare codice offensive
% Capire se con i modelli e' possibile generare attacchi su PowerShell
% Prendiamo un architettura rappresentativa del sota, facendo pre-training e fine-tuning (spiegare nella metodologia)
% Parser --> chiedere dettagli a christian
% Figura --> victor the artist
% Abbiamo creato un dataset, utilzzato per pre-training e fine-tuning (vedi sect. ..)

% RQ: Possiamo generare PowerShell?
% RQ: Pre-training aiuta?
% RQ: Meglio Public (ChatGPT 3.5) o Privato?

% tool estrazione

\begin{figure*}[ht]
\centering
\includegraphics[width=1\linewidth]{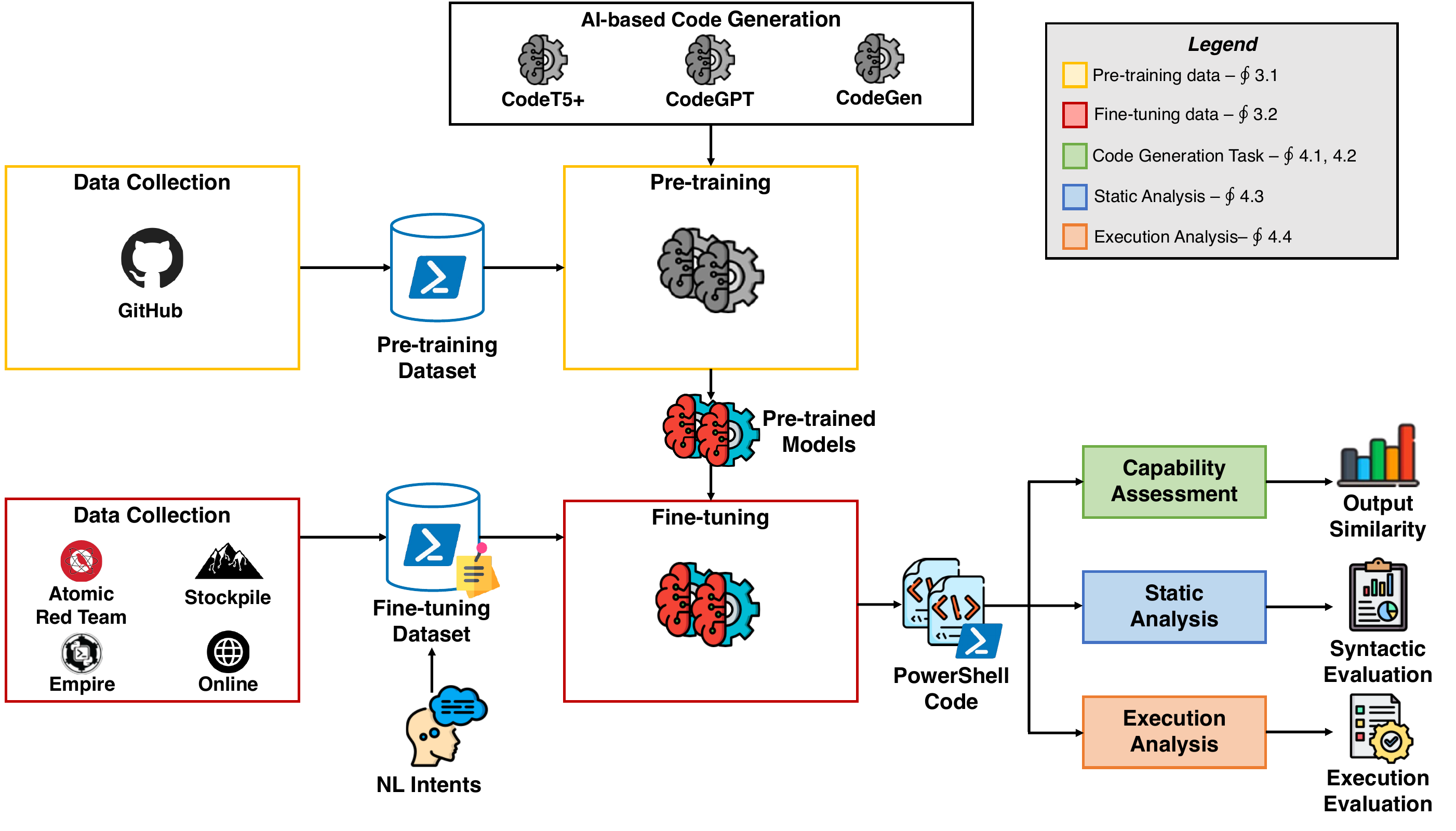}
  \caption{Overview of our research study.}
  \label{fig:approach}
\end{figure*}

The main objective of our research study is to understand whether NMT models can translate NL descriptions into code that accurately replicates the complexities of cyber attacks in PowerShell. 
This aspect is crucial as it explores the models' understanding of the unique syntax and semantics of this programming language. 

\figurename{}~\ref{fig:approach} provides an overview of this research study. We analyze various deep learning strategies to accurately generate code and introduce datasets to train and evaluate them. We study several state-of-the-art NMT models and introduce various approaches to evaluating the generated code, including the similarity of the generated code to ground truth and static and dynamic analysis of the code. 

%\roberto{Analizzeremo varie strategie di deep learning per generare accuratamente il codice, e introdurremo dei dataset ai fini di addestrare e valutare queste strategie. Studieremo vari modelli NMT allo stato dell'arte, e introdurremo vari approcci di valutazione del codice generato, che includono la similarità del codice generato con quello della ground-truth, e l'analisi statica e dinamica del codice.}

%\roberto{Il problema principale per utilizzare i modelli NMT è quello di disporre di un insieme sufficiente di dati, ed utilizzarli efficacemente per addestrare un modello. Ai fini di ottenere sufficienti dati su PowerShell, in questo studio raccoglieremo un ampio insieme di programmi PowerShell utilizzati ai fini di penetration testing e adversary emulation. Oltre al codice, creeremo delle descrizioni di questi programmi in lingua inglese, per poter addrestrare il modello a tradurre l'inglese in codice PowerShell. La creazione di questo dataset è stata svolta manualmente per verificare che i programmi fossero effettivamente legati alla sicurezza, e per garantire che le descrizioni in lingua inglese siano complete e coerenti rispetto al codice. Il dataset è labeled, poiché ogni campione include sia il testo da tradurre in codice, sia il codice che ci si attende venga prodotto dal modello (ground truth). Questo dataset sarà utilizzato per addestrare i modelli NMT mediante fine-tuning, che effettua ulteriori round di addestramento supervisionato del modello per specializzarlo ai fini di generare codice PowerShell.}

To help NMT models in the novel and ambitious task of generating PowerShell code from NL, we adopt a two-step process consisting of \textbf{pre-training} and \textbf{fine-tuning}. The pre-training phase aims to tailor NMT models (already pre-trained on other programming languages) in the generation of PowerShell code. 
%The pre-training phase leverages a substantial dataset extracted from online repositories, including platforms like Github. This dataset comprises a diverse array of general-purpose PowerShell code, serving as a foundational knowledge base to enhance the models' comprehension of PowerShell syntax and semantics. This strategic approach enables the models to grasp the broader context of PowerShell, laying the groundwork for subsequent fine-tuning.
Armed with the pre-trained models, we proceed to the fine-tuning phase. This iterative process refines the models' capabilities, enabling them to generate offensive PowerShell code from NL descriptions. %This approach emphasizes the importance of specialized training data in enhancing the models' performance in the specific domain of offensive security applications.

The main problem in using NMT models is to have a sufficient set of data and to use them effectively to train the models themselves. Recognizing the lack of suitable datasets for offensive PowerShell code generation, in this study, we collect a large set of PowerShell programs used for penetration testing and adversary emulation. In addition to the code, we create descriptions of these programs in English to allow the model to translate English into PowerShell code. This dataset was created manually to verify that the programs were related to security and to ensure that the English language descriptions were complete and consistent with the code. The dataset is labeled since each sample includes both the text to translate into code and the code expected to be produced by the model (ground truth). %This dataset is used to train NMT models through fine-tuning, which performs additional rounds of supervised training to specialize it to generate PowerShell code.

The creation of labeled datasets is inevitably limited by the availability of PowerShell security programs and the need to manually create English language descriptions for each program. To increase the amount of training data, in this study, we investigate an additional strategy, fully automated, to build an extended dataset of PowerShell programs, collecting PowerShell programs and the related text from the web (for example, comments in the code or description accompanying the code). As the collection is fully automated, this second dataset is non-labeled. The dataset includes programs not strictly related to security but includes, in general, PowerShell code used for various purposes. This dataset still contributes to the ability to generate security code since it allows the model to learn from further examples how to generate syntactically valid PowerShell code and to correlate the PowerShell code with the English language. We use this dataset to pre-train the NMT models, carrying out additional unsupervised training rounds.

%We constructed two novel datasets to explore the feasibility of offensive PowerShell code generation through NMT: a \textit{pre-training} and a \textit{fine-tuning} dataset. 
\tablename~\ref{tab:dataset_statistics} reports the statistics of both datasets, in terms of size, unique number of tokens, and average number of tokens for NL descriptions (only for fine-tuning data) and code. %We provide both datasets as open data to encourage further experimentation on this topic. 

%\pietro{Per victor: dire che pubblicheremo il dataset su Github. E che il dataset e' condiviso anonimenete sul link fisghare.}

Finally, we evaluate the models as follows:
\begin{itemize}
    \item \textit{Capability Assessment}: We compare the textual similarity of the code generated by the models with a ground-truth reference through automatic metrics. These metrics are an appealing solution to estimate the generated code since they are easy to tune and time-saving, hence overcoming the limit of human evaluation, which poses practical challenges for large-scale assessments.
    \item \textit{Static analysis}: We assess the generated code to ensure that it adheres to PowerShell programming conventions and does not contain syntax errors.
    \item \textit{Execution analysis}: We evaluate the capability of the generated offensive PowerShell code in executing malicious actions, replicating the behavior of the ground truth commands.
\end{itemize}

%\roberto{Aggiornare la frase precedente per dire che facciamo varie valutazioni (similarità, analisi statica, analisi dinamica).}

In the following of this section, we detail the pre-training (\S{}~\ref{subsec:pretraining}) and the fine-tuning data (\S{}~\ref{subsec:finetuning}), and the code generation task (\S{}~\ref{subsec:cgt}).

\subsection{Pre-training data (unlabeled)}
\label{subsec:pretraining}
%\pietro{dire del MLM}
%\roberto{vedi sezione 3.3.1 della tesi per MLM}
%\roberto{possiamo dire qualche a proposito di domain-adaptive pretraining - https://arxiv.org/pdf/2004.10964.pdf - che si aggiunge al pretraining che i modelli hanno già ricevuto}

Pre-training involves training the model on a large corpus of text data to learn general language representations before fine-tuning it for specific downstream tasks \cite{dai2015semi}. In other words, the parameters obtained from this step serve as a starting point for the later supervised training. Unsupervised or self-supervised pre-training is particularly attractive in the NMT context since large unlabeled data is available on the Internet. In this work, we leverage domain-adaptive pre-training (DAPT) \cite{gururangan2020don}: given an NMT model pre-trained on massive, heterogeneous corpora, we perform additional rounds of unsupervised training with domain-specific data. Specifically, we leverage general-purpose PowerShell code for pre-training. The pre-training dataset aims to provide a valuable resource to enable the models' understanding of general-purpose PowerShell code. This dataset encompasses $\sim90k$ samples extracted through the GitHub API. Specifically, we queried all the repositories containing PowerShell code from the last decade (2013-2023) to encompass a broad spectrum of PowerShell code, then parsed the extracted data to remove unnecessary information, such as duplicates (inside the same repository), and logging and echo commands. In addition, we filtered out all the PowerShell commands with sizes greater than $1024$, ensuring the dataset maintains a balanced representation of code complexities. This collection encompasses a diverse array of PowerShell scripts, spanning various application domains such as system administration, automation, and network management. Including a wide range of scripts reflects the versatility of PowerShell as a scripting language and provides models with exposure to the diverse ways PowerShell is used across different use cases.

The pre-training process depends on the model architecture. For decoder-only models, i.e., CodeGPT and CodeGen, we chose \textit{Causal Language Modeling (CLM)}, also referred to as Language Modeling, as the pre-training objective. CLM has been extensively used as a pre-training task for transformer-based decoder-only models \cite{lin2022survey}, such as in the GPT series \cite{radford2018improving,radford2019language,brown2020language}. CLM refers to language models that predict the next token or sequence of tokens in a sentence in a causal or autoregressive manner, where the prediction for each token depends only on the preceding tokens. By using masking, the model only attends to the left context in a unidirectional manner, ensuring that it cannot see "into the future". In the probabilistic framework, starting from the text sequence $x = (x_1, x_2, x_3, \ldots, x_T)$, where $x$ is the original sentence and $x_t$ ($t = 1, 2, \ldots, T$) is the $t$-th token, and $T$ is the sequence length, an autoregressive model factorizes the likelihood of the input text sequence as \( p(x) = \prod_{t=1}^{T} p(x_t \mid x_{<t}), \) where $p$ is the likelihood of the input text sequence \cite{wang2022pre}. Finally, models are evaluated by token-level accuracy.
For CodeT5+, the pre-training objective is \textit{Masked Language Modeling (MLM)}, as recent works show its effectiveness in code understanding tasks \cite{tufano2023automating}. MLM refers to the prediction of missing tokens in a sentence based on the context provided by the surrounding tokens. Unlike the left-to-right language model pre-training, MLM considers both the left and right context. The approach is inspired by BERT \cite{devlin2018bert}, where 15\% of the tokens in the encoder inputs are randomly replaced with sentinel token [MASK], and the decoder is tasked with recovering these tokens to reconstruct the complete snippet. The model is evaluated by token level accuracy only on the masked-out tokens.
%The pre-training process for CodeGPT and CodeGen is achieved through a slightly modified version of the CodeXGLUE framework \cite{codexglue,text-to-code} for the Code Completion task. While CodeT5+ undergoes its pre-training process using the Hugging Face Transformers library. \\
%As previously explained, all the models were already pre-trained on previous programming datasets. In this work, the models follow a second phase of pre-training, \textit{\textbf{domain-adaptive pre-training}}, as it leads to performance gains in literature \cite{pretrain2, impact_pretraining}. \\
%All the models, including the domain-adapted version, are available on Huggingface. 

% Codice eterogeneo

\begin{table}[t]
\centering
\small
\begin{tabular}{
>{\centering\arraybackslash}m{3.7cm} |
>{\centering\arraybackslash}m{1.8cm}
>{\centering\arraybackslash}m{1.8cm}}
\toprule
\textbf{Statistic} & \textbf{Pre-training Dataset} & \textbf{Fine-tuning Dataset}\\ 
\midrule
\textit{Dataset size} & $89,814$ & $1,127$\\ 
\textit{Unique Intents} & - & $1,077$\\
\textit{Unique Commands} & $79,410$ & $1,121$\\
\textit{Unique tokens (Intents)} & - & $2,273$\\
\textit{Unique tokens (Commands)} & $85,342$ & $17,463$\\ 
\textit{Avg. tokens per Intent} & - & $15.97$\\ 
\textit{Avg. tokens per Command} & $12.71$ & $15.49$\\ 
\bottomrule
\end{tabular}
\caption{Statistics of the pre-training and fine-tuning datasets. The pre-training dataset does not contain NL descriptions (intents).}
\label{tab:dataset_statistics}
\end{table}

\subsection{Fine-tuning data (labeled)}
\label{subsec:finetuning}
%\vittorio{Enfatizzare che il dataset è costruito per essere orientato alla security. La parte labeled contiene solo campioni orientati a security. Controllare la tesi di Christian per informazioni aggiuntive sulle altre fonti di codice (vari siti e blog) e per il mapping con MITRE ATT\&CK.}
The overarching purpose of this dataset is to serve as a comprehensive resource for training models in the translation of NL intents, i.e., descriptions of code snippets, into executable security-oriented PowerShell commands. Specifically, we focus on offensive PowerShell code, a key resource for cybersecurity exercises since Microsoft Windows represents the most targeted OS. By encompassing a wide array of sources, the dataset aims to expose models to the intricacies of real-world cybersecurity scenarios, enabling them to understand and generate PowerShell commands that align with those typical of cybersecurity operations. This holistic approach strives to ensure that models trained on this dataset are well-equipped to handle the complexities of real-world tasks and contribute meaningfully to offensive code generation, specifically PowerShell commands.

The dataset, consisting of $1,127$ samples of PowerShell commands, is meticulously curated from the following sources: 
\begin{itemize}
    \item \textit{Atomic Red Team} \cite{AtomicRedTeam}: renowned for its library of tests mapped to the MITRE ATT\&CK framework\footnote{The ATT\&CK framework is a comprehensive knowledge base of the tactics, techniques, and procedures (TTPs) that adversaries leverage during cyberattacks, developed by MITRE.} \cite{attack}, serves the purpose of replicating real-world adversarial tactics, techniques, and procedures (TTPs). This inclusion provides the dataset with a foundation rooted in a standardized and widely accepted framework, ensuring that the PowerShell commands align with recognized cybersecurity methodologies.
    \item \textit{Stockpile} \cite{Stockpile}: is a plugin for the CALDERA cybersecurity framework \cite{Applebaum2016IntelligentAR, CALDERAGitHub} developed by MITRE and introduces a layer of sophistication by incorporating structured data integral for adversary emulation. Therefore, the dataset does not encompass raw PowerShell commands only but also captures the contextual information and relationships between commands within the broader context of adversarial scenarios.
    \item \textit{Empire} \cite{Empire}: a post-exploitation and adversary emulation framework integrated with MITRE ATT\&CK, provides PowerShell commands representative of advanced malicious techniques, further enriching the dataset with nuanced and intricate scenarios.
%    \item \textit{ChatGPT API}: including PowerShell commands generated through interactions with the ChatGPT API from OpenAI introduces a dynamic and interactive element to the dataset. This source enables the incorporation of NL descriptions and intents into the dataset, reflecting real-world scenarios where users may describe a desired action in NL before translating it into a PowerShell command. \roberto{Possiamo dire qualcosa in più su che tipo di prompt abbiamo usato?}
%    \pietro{Riporto una frase di un reviewer su un paper simile: ChatGPT assessment: the description of the experiment involving ChatGPT is too vague. The authors should provide the exact prompts provided to ChatGPT, ensure to perform the assessment in different conversations to avoid leakage and perform this assessment multiple times for each sample given the stochastic nature of the tool.}
%    \pietro{discutere oggi}
    \item \textit{Online sources}: we manually verified and selected additional offensive samples from several security-related online sources. We gathered samples from \textit{HackTricks} \cite{hacktricks}, \textit{Red Team Recipe} \cite{pwshpentest}, and \textit{Infosec Matter} \cite{blogpentest}, community-driven cybersecurity wikis about ethical hacking, penetration testing, and information security. By including diverse examples specific to the offensive PowerShell dataset, the model acquires a more profound understanding of the conventions and best practices unique to PowerShell security commands.
\end{itemize}

\begin{figure}
    \centering
    \includegraphics[width=1\columnwidth]{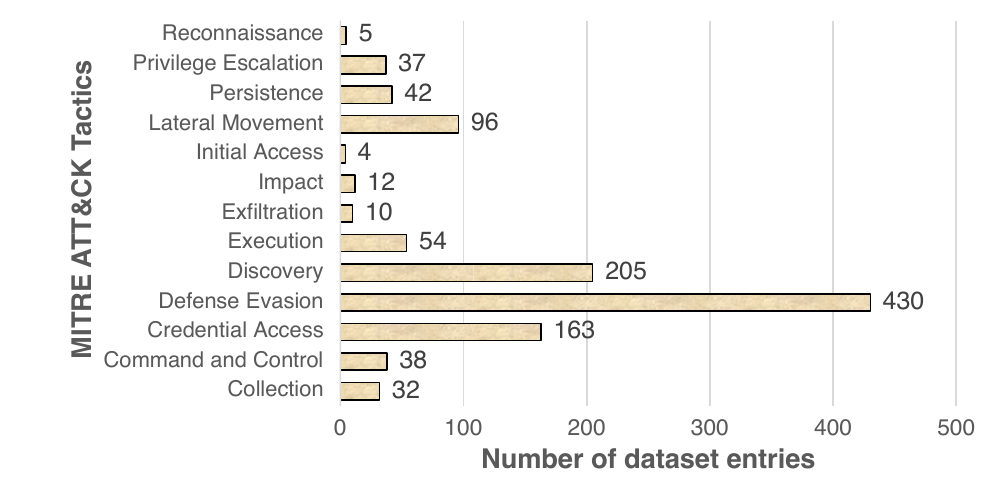}
    \caption{Mapping of fine-tuning dataset samples on the MITRE ATT\&CK tactics.}
    \label{fig:Mapping fine-tuning ATT&CK}
\end{figure}

We manually curated the dataset to cover the highest number of tactics in the MITRE ATT\&CK framework. In particular, the dataset covers 12 out of 14 tactics from the MITRE ATT\&CK framework, the \textit{de facto} standard for adversarial techniques representation, with varying numbers of techniques and sub-techniques per tactic. Figure \ref{fig:Mapping fine-tuning ATT&CK} illustrates the number of entries for each ATT\&CK tactic. Each entry in the dataset is annotated with an NL description extracted from the respective source. We manually annotated every sample that did not come with a predefined description. Moreover, we enriched all those descriptions that did not provide enough information about the specific PowerShell command. For instance, in the case of Atomic Red Team, the PowerShell commands represent implementations of the techniques in the ATT\&CK framework. Consequently, these commands are often labeled with the technique name, which provides informative content about the technique itself rather than what the command does. To better understand how programmers and security experts describe PowerShell scripts and how to deal with ambiguities in natural language, we referred to popular books and manuals \cite{TutorialsPointPowerShell, lee2011windows, holmes2012windows}.

Finally, we notice that the size of our dataset is in line with other state-of-the-art corpora used to fine-tune ML models. In fact, in state-of-the-art code generation, the datasets for fine-tuning are relatively limited, in the order of one thousand samples~\cite{zhou2023lima}.

\begin{comment}

\begin{table}[ht]
\centering
\caption{Mapping of fine-tuning dataset samples on the MITRE ATT\&CK tactics. \textbf{\textcolor{red}{Sostituire con un istogramma che rappresenta solamente il numero di campioni per ogni tattica}}}
\begin{center}
\resizebox{\columnwidth}{!}{%
\begin{tabular}{cccccc}
\toprule
    \textbf{Tactic} & \textbf{ChatGPT} & \textbf{Atomic} & \textbf{Empire} & \textbf{Stockpile} & \textbf{Online} \\
    \midrule
    Collection & 0 & 14 & 5 & 7 & 6 \\
    \midrule
    Command \& Control & 1 & 24 & 6 & 0 & 7 \\
    \midrule
    Credential Access & 1 & 41 & 66 & 4 & 50 \\
    \midrule
    Defense Evasion & 244 & 95 & 33 & 7 & 51 \\
    \midrule
    Discovery & 0 & 42 & 29 & 41 & 93 \\
    \midrule
    Execution & 11 & 13 & 4 & 5 & 21 \\
    \midrule
    Exfiltration & 0 & 8 & 1 & 0 & 1 \\
    \midrule
    Impact & 0 & 5 & 0 & 6 & 1 \\
    \midrule
    Initial Access & 2 & 0 & 0 & 0 & 2 \\
    \midrule
    Lateral Movement & 62 & 6 & 9 & 5 & 14 \\
    \midrule
    Persistence & 1 & 15 & 15 & 0 & 11 \\
    \midrule
    Privilege Escalation & 0 & 7 & 11 & 2 & 17 \\
    \midrule
    Reconnaissance & 1 & 2 & 0 & 0 & 2 \\
    \midrule
    \textbf{Total} & 323 & 272 & 179 & 77 & 276 \\
    \bottomrule
\end{tabular}}%
\label{tab:Mapping fine-tuning ATT&CK}
\end{center}
\end{table}

\end{comment}

% dire che abbiamo lasciato i commenti, ove disponibili
% descrivere annotazione manuale: i commenti "poveri" a livello di contenuto informativo sono stati arricchiti dagli autori in base alla loro conoscenza

\subsection{Code Generation Task}
\label{subsec:cgt}

To ensure the robustness of our study, we adopt the following state-of-the-art NMT models:
\begin{itemize}
    \item \textbf{CodeT5+}~\cite{wang2023codet5+} is a new family of Transformer models pre-trained with a diverse set of pretraining tasks to learn rich representations from both unimodal code data and bimodal code-text data. We utilize the variant with model size $220M$, trained from scratch following T5’s architecture~\cite{DBLP:journals/jmlr/RaffelSRLNMZLL20}. It has an encoder-decoder architecture with $12$ decoder layers, each with $12$ attention heads and hidden layer dimension of $768$, and $512$ for the size of position embeddings. We set the learning rate $\alpha = 0.00005$, batch size = $16$, and beam size = $10$.
    \item \textbf{CodeGPT}~\cite{lu2021codexglue}, a Transformer-based language model pre-trained on millions of Python functions and Java methods. The model architecture consists of 12 layers of Transformer decoders. We followed previous work for the implementation~\cite{CodeXGLUE}.
    \item \textbf{CodeGen} \cite{codegen}, an autoregressive language model for program synthesis with an architecture that follows a standard transformer decoder with left-to-right causal masking. The family of CodeGen models is trained in various sizes, including 350M, 2.7B, 6.1B, and 16.1B, and utilizes various datasets. Specifically, we leverage CodeGen-Multi, initialized from CodeGen-NL and further pre-trained on BigQuery \cite{codegen},  a large-scale dataset of multiple programming languages from GitHub repositories, which consists of 119.2B tokens and includes C, C++, Go, Java, JavaScript, and Python.
    
\end{itemize}

In our experiments, we randomly split the fine-tuning dataset into training (the set of examples used to fit the parameters), validation (the set used to tune the hyperparameters of the models), and test (the set used for the evaluation of the models) sets using a typical $80\%/10\%/10\%$ ratio.

%\roberto{per coerenza, dovremmo spostare tutte le metriche qui (anche quelle per l'analisi statica e dinamica), oppure spostare le metriche di similarity nella sezione che le analizza}

To assess the performance of the models in generating offensive PowerShell code from NL descriptions, we used \emph{output similarity metrics}, which compare the generated code with the code from the ground truth. This type of metrics is widely used to assess the performance of AI generators in many code generation tasks~\cite{LIGUORI2023120073}, including the generation of code for security contexts~\cite{yang2022dualsc,yang2023exploitgen,ruan2023prompt,liguori2021evil,liguori2022can}. The metrics are: 

\begin{itemize}
    \item \textbf{Bilingual Evaluation Understudy (BLEU) score}~\cite{papineni2002bleu}. It measures the degree of n-gram overlapping between the string of each code snippet produced by the model and the reference, for values of $n$ usually ranging between $1$ and $4$ \cite{han2016machine,munkova2020evaluation}. 
    %This metric also takes into account a \textit{brevity penalty} to penalize predictions shorter than the references. BLEU value ranges between $0$ and $1$, with higher scores corresponding to a better quality of the prediction. 
    We implemented BLEU-4 score (i.e., with $n=4$) computation employing the \texttt{bleu\_score} module contained in the open-source Natural Language Toolkit (NLTK) Python suite~\cite{bleu}.

    \item \textbf{Edit Distance (ED)}. It measures the \textit{edit distance} between two strings, i.e., the minimum number of operations on single characters required to make each code snippet produced by the model equal to the reference. For the edit distance, we adopted the Python library \texttt{pylcs}~\cite{pylcs}. 
    
    \item \textbf{METEOR}\cite{10.5555/1626355.1626389}. It measures the \textit{alignment} between each code snippet produced by the model and the reference. The alignment is defined as a mapping between unigrams (i.e., $1$-gram), such that every unigram in each string maps to zero or one unigram in the other string and no unigrams in the same string. To calculate the METEOR metric, we relied on the Python library \texttt{evaluate} by HuggingFace~\cite{meteor}.
    
    \item \textbf{ROUGE-L}. It is a metric based on the longest common subsequence (LCS) between the model output and the reference, i.e., the longest sequence of words (not necessarily consecutive, but still in order) shared between both. We computed the ROUGE-L metric using the Python package \texttt{rouge}~\cite{rouge}.
\end{itemize}

All metrics range between $0$ and $1$, with higher scores corresponding to a better quality of the generated code. To evaluate the generated PowerShell code, we also introduce additional evaluation metrics based on static and dynamic analysis that are specific to our context. These metrics will be introduced in the following sections.

\subsection{Research Questions}
\label{subsec:questions}

We designed this research study to answer the following research questions (RQs):

\noindent
$\rhd$ \textbf{RQ1:} \textit{To what extent can NMT models effectively generate offensive PowerShell code for security applications from NL descriptions?}\\
RQ1 aims to establish a preliminary assessment of NMT models in generating PowerShell code for offensive security applications. This investigation seeks to shed light on the models' efficacy in translating NL descriptions into offensive code.

\noindent
$\rhd$ \textbf{RQ2:} \textit{What is the influence of the training strategies on NMT models' performance in offensive PowerShell code generation?}\\
RQ2 focuses on the impact of pre-training and fine-tuning on the quality of generated code. We analyze the influence of these training strategies by considering different configurations of the NMT models and their impact on their performance.

\noindent
$\rhd$ \textbf{RQ3:} \textit{How good is the generated code in terms of code quality and dynamic behavior?}\\
RQ3 aims to evaluate the generated PowerShell code in a deeper way than output similarity metrics, in terms of syntactic correctness and capability of executing malicious actions realistically, through behavioral comparison with the ground truth.
%\pietro{Aggiungere descrizione RQ, poi ci focalizziamo sulla domanda}

\noindent
$\rhd$ \textbf{RQ4:} \textit{How do fine-tuned NMT models, leveraging security-oriented training data, compared to a publicly available, closed-source model?}\\
RQ4 introduces a comparative analysis, evaluating the performance of the fine-tuned models against a publicly available general-purpose language model, specifically ChatGPT 3.5. This investigation strives to evaluate whether specialization on security-focused data provides an advantage in the offensive PowerShell code generation domain.
\section{Experimental Results}
\label{sec:evaluation}
This section presents an extensive evaluation of NMT models (CodeT5+, CodeGPT, and CodeGen) on the generation of offensive PowerShell code. First, we assess the models' capability of generating PowerShell code in their original configuration (\S{}~\ref{subsec:zero-shot}) without further training. Then, we evaluate the impact of different training strategies, i.e., domain-adaptive pre-training and fine-tuning, on the performance of such models (\S{}~\ref{subsec:impact}). To provide further insight into the PowerShell code generation, we analyze the quality of the generated code in terms of syntactic correctness (\S{}~\ref{subsec:static}) and dynamic behavior (\S{}~\ref{subsec:execution}), i.e., its ability to replicate the behavior of the ground truth code. Finally, we compare the fine-tuned models with a public AI model (ChatGPT) for all the previous analyses (\S{}~\ref{subsec:comparison}) to benchmark their performance against a publicly available, closed-source model.

%\vittorio{Fornire più informazioni sulla specifica tecnica di pre-training, per valorizzare questa parte. \\ Aggiornare i risultati con il terzo modello, e con la parte di analisi dinamica}

%\vittorio{Aggiungere preambolo a questa sezione (ora che le RQ sono in sezione 3) magari spiegando quali sono le analisi presentate in questa sezione e com'è suddivisa la sezione stessa}

\subsection{Zero-shot Learning}
\label{subsec:zero-shot}
To establish a baseline for the evaluation, we initially used the NMT models in their original configuration,  asking them to generate PowerShell code. This is a \textit{zero-shot learning task}, where an NMT model is applied for a different scenario than the one for which it was trained. In this way, we evaluate the current gap of existing models in generating PowerShell code. Table \ref{tab:results-zeroshot} shows the results of this analysis. In this task, the models are tested without any gradient updates, relying only on the intent provided by the test set for inference \cite{radford2019language, brown2020language}. 
The non-pre-trained versions of the models tend to generate Python code, but their performance is generally low for the downstream task of generating offensive PowerShell code. Pre-training the models with general-purpose PowerShell code slightly improves the accuracy but is still not high. Among the pre-trained versions, CodeGPT is the only one that provides output close to valid PowerShell code, although it does not align well with the expected code indicated by the intent in natural language. 
In summary, regardless of pre-training, all models demonstrate the need for fine-tuning on a tailored dataset for optimal performance in generating offensive PowerShell code.

\begin{table}[t]
\centering
\footnotesize
\begin{tabular}{
>{\centering\arraybackslash}m{1.1cm} |
>{\centering\arraybackslash}m{0.8cm} |
>{\centering\arraybackslash}m{1cm}
>{\centering\arraybackslash}m{1.1cm} 
>{\centering\arraybackslash}m{1cm}
>{\centering\arraybackslash}m{1cm} } 
\toprule
\textbf{Model}  & \textbf{Pre-training} & \textbf{BLEU-4 (\%)} & \textbf{ED (\%)} & \textbf{METEOR (\%)} &\textbf{ROUGE-L (\%)}\\
\midrule
\multirow{2}{*}{\textbf{CodeT5+}} 
    & \xmark & 0.04 \DrawPercentageBar{0.0004} & 8.87 \DrawPercentageBar{0.0887} & 4.69 \DrawPercentageBar{0.0469} & 1.08 \DrawPercentageBar{0.0108} \\
    & \cmark & 0.01 \DrawPercentageBar{0.0001} & 6.96 \DrawPercentageBar{0.0696} & 1.86 \DrawPercentageBar{0.0186} & 2.68 \DrawPercentageBar{0.0268} \\
\midrule
\multirow{2}{*}{\textbf{CodeGPT}}
    & \xmark & 0.23 \DrawPercentageBar{0.00023} & 12.31 \DrawPercentageBar{0.1231} & 4.08 \DrawPercentageBar{0.0408} & 1.19 \DrawPercentageBar{0.0119} \\
    & \cmark & 0.28 \DrawPercentageBar{0.00028} & 15.67 \DrawPercentageBar{0.1568} & 2.55 \DrawPercentageBar{0.0255} & 3.41 \DrawPercentageBar{0.0341} \\
\midrule
\multirow{2}{*}{\textbf{CodeGen}}
    & \xmark & 0.06 \DrawPercentageBar{0.0006} & 7.58 \DrawPercentageBar{0.0758} & 2.88 \DrawPercentageBar{0.0288} & 0.21 \DrawPercentageBar{0.00021} \\
    & \cmark & 0.00 \DrawPercentageBar{0} & 0.43 \DrawPercentageBar{0.00043} & 0.09 \DrawPercentageBar{0.0009} & 0.00 \DrawPercentageBar{0} \\
\bottomrule
\end{tabular}%
\caption{Performance of models with and without pre-training on zero-shot.}
\label{tab:results-zeroshot}
\end{table}

%\roberto{Possiamo aggiungere nel testo quali sono i valori "tipici" di BLEU, etc. in altri contesti? Ad esempio i valori riportati negli articoli di questi paper, o altri articoli che valutano il codice. Per dire che questi modelli hanno una qualità sotto lo standard. Possiamo fare questo confronto nel testo anche nella sezione successiva.}

\begin{table}[t]
\renewcommand{\arraystretch}{1.2}
\centering
%\makebox[\textwidth]{% 
\centering
\scriptsize
\begin{tabular}%{lcccccccc}
{
>{\centering\arraybackslash}m{0.9cm} |
>{\centering\arraybackslash}m{0.5cm}
>{\centering\arraybackslash}m{0.5cm}
>{\arraybackslash}m{0.93cm}
>{\arraybackslash}m{0.93cm}
>{\arraybackslash}m{0.93cm}
>{\arraybackslash}m{0.93cm}} 
\toprule
\textbf{Model} & \textbf{Epochs} & \textbf{Pre-train. (\%)} & \textbf{BLEU-4 (\%)} & \textbf{ED (\%)} & \textbf{METEOR (\%)} &\textbf{ROUGE-L (\%)}\\
\midrule
\multirow{7}{*}{\textbf{CodeT5+}} 
    & \multirow{2}{*}{3} & \xmark & 4.22 \DrawPercentageBar{0.0422} & 35.11 \DrawPercentageBar{0.3511} & 28.83 \DrawPercentageBar{0.2883} & 22.26 \DrawPercentageBar{0.2226} \\
    & & \cmark & 4.57 \DrawPercentageBar{0.0457}  & 35.96 \DrawPercentageBar{0.3596} & 30.57 \DrawPercentageBar{0.3057} & 23.99 \DrawPercentageBar{0.2399} \\
\cmidrule{2-7}
    & \multirow{2}{*}{10} & \xmark & 12.64 \DrawPercentageBar{0.1264} & 46.72 \DrawPercentageBar{0.4672} & 44.76 \DrawPercentageBar{0.4476} & 37.65 \DrawPercentageBar{0.3765} \\
    & & \cmark & 11.88 \DrawPercentageBar{0.1188} & 49.10 \DrawPercentageBar{0.491} & 46.11 \DrawPercentageBar{0.4611} & 37.17 \DrawPercentageBar{0.3717} \\
\cmidrule{2-7}
    & \multirow{2}{*}{30} & \xmark & 17.40 \DrawPercentageBar{0.174} & \textcolor{blue}{\textbf{50.92}} \DrawPercentageBarBlue{0.5092} & 47.61 \DrawPercentageBar{0.4761} & \textcolor{blue}{\textbf{39.05}} \DrawPercentageBarBlue{0.3905} \\
    & & \cmark & 18.50 \DrawPercentageBar{0.185} & 50.23 \DrawPercentageBar{0.5023} & \textcolor{blue}{\textbf{47.87}} \DrawPercentageBarBlue{0.4787} & 38.86 \DrawPercentageBar{0.3886} \\
\midrule

\multirow{7}{*}{\textbf{CodeGPT}}
    & \multirow{2}{*}{3} & \xmark & 10.28 \DrawPercentageBar{0.1028} & 40.71 \DrawPercentageBar{0.4071} & 31.21 \DrawPercentageBar{0.3121} & 25.60 \DrawPercentageBar{0.256} \\
    & & \cmark & 12.80 \DrawPercentageBar{0.128} & 42.54 \DrawPercentageBar{0.4254} & 35.14 \DrawPercentageBar{0.3514} & 30.35 \DrawPercentageBar{0.3035} \\
\cmidrule{2-7}
    & \multirow{2}{*}{10} & \xmark & 16.22 \DrawPercentageBar{0.1622} & 46.39 \DrawPercentageBar{0.4639} & 40.50 \DrawPercentageBar{0.405} & 33.52 \DrawPercentageBar{0.3352} \\
    & & \cmark & 17.93  \DrawPercentageBar{0.1793} & 49.88 \DrawPercentageBar{0.4988} & 45.12 \DrawPercentageBar{0.4512} & 37.12 \DrawPercentageBar{0.3712} \\
\cmidrule{2-7}
    & \multirow{2}{*}{30} & \xmark & \textcolor{blue}{\textbf{21.71}} \DrawPercentageBarBlue{0.2171} & 50.17 \DrawPercentageBar{0.5017} & 45.34 \DrawPercentageBar{0.4534} & 38.63 \DrawPercentageBar{0.3863} \\
    & & \cmark & 19.94 \DrawPercentageBar{0.1994} & 49.20 \DrawPercentageBar{0.492} & 45.45 \DrawPercentageBar{0.4545} & 38.06 \DrawPercentageBar{0.3806} \\
\midrule

\multirow{7}{*}{\textbf{CodeGen}}
    & \multirow{2}{*}{3} & \xmark & 16.20 \DrawPercentageBar{0.162} & 47.68 \DrawPercentageBar{0.4768} & 42.27 \DrawPercentageBar{0.4227} & 35.97 \DrawPercentageBar{0.3597} \\
    & & \cmark & 14.75 \DrawPercentageBar{0.1475} & 45.88 \DrawPercentageBar{0.4588} & 39.86 \DrawPercentageBar{0.3986} & 34.69 \DrawPercentageBar{0.3469} \\
\cmidrule{2-7}
    & \multirow{2}{*}{10} & \xmark & 19.15 \DrawPercentageBar{0.1915} & 50.52 \DrawPercentageBar{0.5052} & 46.76 \DrawPercentageBar{0.4676} & 37.63 \DrawPercentageBar{0.3763} \\
    & & \cmark & 19.04 \DrawPercentageBar{0.1904} & 48.45 \DrawPercentageBar{0.4845} & 43.25 \DrawPercentageBar{0.4325} & 35.25 \DrawPercentageBar{0.3525} \\
\cmidrule{2-7}
    & \multirow{2}{*}{30} & \xmark & 18.23 \DrawPercentageBar{0.1823} & 47.53 \DrawPercentageBar{0.4753} & 44.10 \DrawPercentageBar{0.441} & 35.48 \DrawPercentageBar{0.3548} \\
    & & \cmark & 18.53 \DrawPercentageBar{0.1853} & 48.67 \DrawPercentageBar{0.4867} & 44.14 \DrawPercentageBar{0.4414} & 35.45 \DrawPercentageBar{0.3545} \\
\bottomrule
\end{tabular}
%}
\caption{Performance of models with and without pre-training and different number of epochs. Best results for each metric are \textcolor{blue}{\textbf{blue/bold}}.}
\label{tab:results}
\end{table}

\subsection{Impact of Training Strategies}
\label{subsec:impact}
The evaluation of CodeT5+, CodeGPT, and CodeGen involved a meticulously designed test plan.
More precisely, the models underwent three distinct fine-tuning scenarios: 3 Epochs, 10 Epochs, and 30 Epochs. This deliberate choice allowed us to assess the impact of prolonged fine-tuning on the models' ability to generate PowerShell code for offensive security tasks. In each scenario, we considered two training configurations: one with pre-training and the other without. This test plan allowed us to systematically explore the models' capabilities under varying conditions, providing a comprehensive understanding of their strengths and limitations. \tablename{}~\ref{tab:results} shows the results.

In the 3 epochs setting, CodeT5+ exhibits low performance, regardless of pre-training, with a BLEU-4 score lower than 10\%. In contrast, CodeGPT and CodeGen demonstrate notable performance even after a short fine-tuning period, achieving a BLEU-4 score higher than 10\% and an ED over 40\%. Notably, after 3 epochs, CodeGen demonstrates superior performance compared to the other two models.
In the 10 epochs experiment, CodeT5+ shows significant improvement, with BLEU-4 tripling to 12\%. Moreover, ED, METEOR, and ROUGE-L experience a rise of 12-16\%. CodeGPT also enhances its performance, surpassing CodeT5+ in terms of BLEU-4 score, although it faces challenges in achieving the same level of overall improvement. CodeGen remains ahead of the other models, even reaching an ED over 50\%.
For a more in-depth assessment of the models' adaptability, the training duration is extended to 30 epochs. CodeT5+ demonstrates superior performance over CodeGPT in ED, METEOR, and ROUGE-L metrics, while CodeGPT exhibits a higher BLEU-4 score surpassing 20\%. Notably, both models achieve a high ED value of around 50\%. CodeGen establishes its performance without further improvement compared to the 10 epochs versions.

%In the 3 Epochs setting, the models performed a relatively brief fine-tuning period, enabling us to observe their initial performance. Even without pre-training, CodeT5+ demonstrated a commendable BLEU-4 score of $0.27$, underscoring its inherent capability. Conversely, CodeGPT struggled to match this performance, with respective BLEU-4, ED, and METEOR of $0.13$, $0.41$, and $0.37$. 

%For a more in-depth assessment, we extended the training duration to 10 Epochs. This choice allowed us to estimate the models' adaptability and performance over an extended fine-tuning period. CodeT5+ continued to dominate performance evaluation, achieving values of BLEU-4, ED, and METEOR equal to $0.38$, $0.51$, and $0.47$, respectively, which represent the best performance. CodeGPT, while displaying improvement, still lagged behind with BLEU-4 scores of $0.18$, ED of $0.48$, and METEOR of $0.44$.

To provide an estimate of the goodness of the results, we compared the results of the models with the performance of the state-of-the-art (SOTA). Since the task of generating PowerShell using NMT models is a task never addressed before, we compared the results with recent work investigating the effectiveness of existing models in the generation of different languages from NL, specifically, Python code~\cite{shin2023good} and in shell language~\cite{shi2023shellgpt}. We found that the best performance is 21\% for BLEU-4 and 38\% for METEOR in the case of the Python language, and 25\% for BLEU-4 and 44\% for ED in the case of shell language. 
We notice that our results are in line with the ones of the SOTA. Even better, our best performance, represented by CodeT5+ without pre-training and 30 fine-tuning epochs, overcomes the SOTA over all the metrics.

We also assessed the impact of varying the number of epochs on fine-tuning time, with distinct differences observed between 3, 10, and 30 epochs for each model.
For both CodeT5+ and CodeGPT, fine-tuning over 3 epochs takes approximately $20$ minutes, whereas CodeGen requires double that time (40 minutes). 
Extending to 10 epochs, CodeT5+ and CodeGPT need around $35$ and $39$ minutes, respectively, while CodeGen's training time increases to $90$ minutes. For the 30-epoch extension, CodeT5+ takes about $80$ minutes, CodeGPT requires $110$ minutes, and CodeGen extends its training time to $270$ minutes.
Finally, the comparison between the fine-tuning times of pre-trained and non-pre-trained models did not reveal evident differences, suggesting that the pre-training process does not introduce a significant computational overhead during the subsequent fine-tuning phase.

%In summary, the fine-tuning time analysis highlights the importance of considering the trade-off between computational resources and model performance when determining the optimal number of training epochs. While increasing epochs may contribute to improved model performance, the associated rise in fine-tuning time should be weighed against the potential benefits. Additionally, the comparable fine-tuning times for pre-trained and non-pre-trained models indicate that the pre-training step does not significantly affect the computational efficiency of the subsequent fine-tuning process for these models.

\begin{mybox}{\parbox{8cm}{RQ1: To what extent can state-of-the-art NMT models effectively generate offensive PowerShell code for security applications from NL descriptions?}}
The evaluation of CodeT5+, CodeGPT, and CodeGen underscores their remarkable effectiveness in generating offensive PowerShell code for security applications from NL descriptions. CodeGen surpasses the other models in the 3 and 10 epochs settings according to all metrics. CodeT5+, designed with a specialized architecture for code generation tasks, consistently outperforms CodeGPT and CodeGen across various metrics in the 30 epochs setting. Particularly noteworthy is the comparison with SOTA performance in code generation tasks for different languages, such as Python and shell languages. Our best-performing model, CodeT5+ without pre-training and 30 fine-tuning epochs, surpasses the SOTA results, exhibiting superior performance across all metrics. 
\end{mybox}

\begin{table*}[t]
\small
\centering
\begin{tabular}{
%>{\centering\arraybackslash}m{1cm} |
 >{\centering\arraybackslash}m{4cm}|
 >{\centering\arraybackslash}m{6.25cm}|
 >{\centering\arraybackslash}m{6.25cm}}
\toprule
%\textbf{Dataset} & \textbf{Natural Language Intent} & \textbf{Ground Truth} & \textbf{Model Output}\\ \midrule
\textbf{Natural Language Intent} & \textbf{Ground Truth} & \textbf{Model Output}\\ \midrule
\textit{Start a process with DirLister, wait for 4 seconds, and stop the DirLister process.} & \texttt{Start-Process \$\{WebBrowserPassViewPath\} ; Start-Sleep -Second 4 ; Stop-Process -Name "WebBrowserPassView"} & \texttt{Start-Process \$\{WebBrowserPassViewPath\} ; Start-Sleep -Second 4 ; Stop-Process -Name "WebBrowserPassView"} \\ \cmidrule(l){1-3} 
\textit{Invoke-Mimikatz cmdlet with bypassed execution policy.} & \texttt{powershell.exe -exec bypass -c `` Invoke-Mimikatz ''} & \texttt{powershell.exe -ExecutionPolicy Bypass -Command `` Invoke-Mimikatz ''} \\ \midrule
\textit{Invoke compiled HTML Help (CHM) file with specific storage handler, HH file path, and CHM file path.} & \texttt{Invoke-ATHCompiledHelp -InfoTechStorageHandler \$ \{ infotech\_storage\_handler \} -HHFilePath \$ \{ hh\_file\_path \} -CHMFilePath \$ \{ chm\_file\_path \}} & \texttt{Invoke-ATHCompiled\textbf{\textcolor{red}{HTML}}Help \textbf{\textcolor{red}{\cancel{-InfoTechStorageHandler }}} \textbf{\textcolor{red}{\cancel{\$ \{ infotech\_storage\_handler \}}}} -HHFilePath \$ \{ hh\_file\_path \} -CHMFilePath \$ \{ chm\_file\_path \}} \\ \bottomrule
\textit{Get the wininit process and perform token manipulation to create a new process for 'cmd.exe'.} & \texttt{Get-Process wininit | Invoke-TokenManipulation -CreateProcess 'cmd.exe'} & \texttt{\textcolor{red}{\textbf{\$wininit = }}Get-Process wininit | Invoke-TokenManipulation -CreateProcess 'cmd.exe'} \\ \bottomrule
\end{tabular}
\caption{Illustrative examples of model output. The prediction errors are \textcolor{red}{\textbf{red/bold}}. \textcolor{red}{\textbf{\cancel{Slashed}}} text refers to omitted predictions.}
\label{tab:model_predictions}
\end{table*}

Considering the impact of pre-training further enriched our evaluation.
Focusing on the 3-epoch experiments, CodeT5+ exhibits a slight improvement across all metrics, and CodeGPT extends the improvement to 2\%-4\% across all metrics. Conversely, CodeGen appears to have better performance without pre-training. Training the models for 10 epochs reveals a more pronounced distinction between the two versions. CodeT5+ pre-training results in a 2\% increase in both Edit Distance (ED) and METEOR metrics. CodeGPT, on the other hand, shows a substantial displacement of 1.7\%, 3.5\%, 4.6\%, and 3.6\% for BLEU-4, ED, METEOR, and ROUGE-L, respectively. CodeGen maintains a negative displacement between the versions even with the extended training duration. When extending the fine-tuning duration to 30 epochs, pre-training did not consistently yield superior results. In this case, the performance of pre-trained models is comparable to non-pre-trained counterparts.

%While CodeT5+ exhibited an improvement across all metrics when subjected to pre-training, this advantage was evident only when the fine-tuning duration was limited to 3 epochs, leading to an increment of the metrics of $0.04$, $0.03$, and $0.05$ for BLEU-4, ED, and METEOR, respectively. Interestingly, pre-training did not consistently yield superior results for extended fine-tuning durations, such as 10 epochs for CodeT5+ and any epoch setting for CodeGPT. In these scenarios, the performance of pre-trained models was lower than that of their non-pre-trained counterparts.

\begin{mybox}{\parbox{8cm}{RQ2: What is the influence of the training strategies on NMT models' performance in offensive PowerShell code generation?}}
As the fine-tuning period extends, such as with 10 and 30 epochs, the benefits of pre-training diminish or even become counterproductive. In these cases, the performance of pre-trained models consistently falls below that of their non-pre-trained counterparts. This highlights the variable effectiveness of pre-training, dependent on the duration of fine-tuning. These findings underscore the interplay between the duration of training epochs and the usage of pre-training, emphasizing the importance of carefully considering these factors in model development. 
\end{mybox}

%\luciano{Rifatto le figure sulla base di quelle della tesi di Christian Ps. questa sezione va lasciata quì o messa sotto (così come nella tesi)? }

\tablename~\ref{tab:model_predictions} illustrates four cases of model predictions. They are examples from our test sets to highlight both successful and failed prediction cases. 
Row \# 1 demonstrates the models' ability to generate a PowerShell snippet composed of multiple commands (separated by semicolons) without errors. The model correctly predicts the correct variables, e.g., \texttt{WebBrowserPassViewPath}, and command names, such as \texttt{Start-Process}, \texttt{Start-Sleep}.
Row \# 2 is indicative of the concept of implicit model knowledge. Indeed, the model can generate a correct command by leveraging alternative equivalent versions of PowerShell's option flags (e.g., \texttt{-ExecutionPolicy} instead of \texttt{-exec}).
Row \# 3 shows a relevant example of a failure case. It is possible to notice how the model correctly predicts the variable names and values except for one not referenced in the intent (\texttt{-InfoTechStorageHandler}). In addition, the model fails to predict the correct command name, generating an additional word (\texttt{HTML}) based on the NL description.
Finally, row \# 4 illustrates another incorrect example in which the model is capable of generating the ground truth code, except for introducing an additional variable to save the output of the command (\texttt{\$wininit =}).

Overall, we can conclude that these examples indicate the model's ability to generate complex PowerShell snippets, even though there is still some error margin, specifically related to omissions (e.g., variable names).

\subsection{Static Analysis}
\label{subsec:static}
We evaluated the generated code through \textit{static analysis} to ensure that the code adheres to PowerShell conventions and does not contain syntax errors. The analysis was conducted on the top-performing models identified in the previous evaluation, namely the 30-epoch versions of CodeT5+ with pre-training, CodeGPT without pre-training, and CodeGen with pre-training. The static analysis leverages \textit{PSScriptAnalyzer} \cite{psscript}, a static code checker for PowerShell modules and scripts. The primary purpose of PSScriptAnalyzer is to assess the quality of PowerShell code by analyzing its syntax, structure, and adherence to best practices. The rules are based on PowerShell best practices identified by the PowerShell Team and the community, organized into categories such as Cmdlet Design, Script Functions, Error Handling, Scripting Style, and Script Security. The severity levels (ParseError, Error, Warning, Information) associated with each rule indicate the importance and impact of adhering to the specific guideline.
In this work, we focused on \textit{parse errors}, which occur during the parsing phase of a program's execution, \textit{errors}, occurring when code does not meet specific high-severity rules (e.g., hardcoding computer names, using plain text passwords), and \textit{warnings}, which typically highlight potential issues or coding practices that might lead to errors or security concerns. 

\begin{figure}[t]
    \centering
    \includegraphics[width=1\columnwidth]{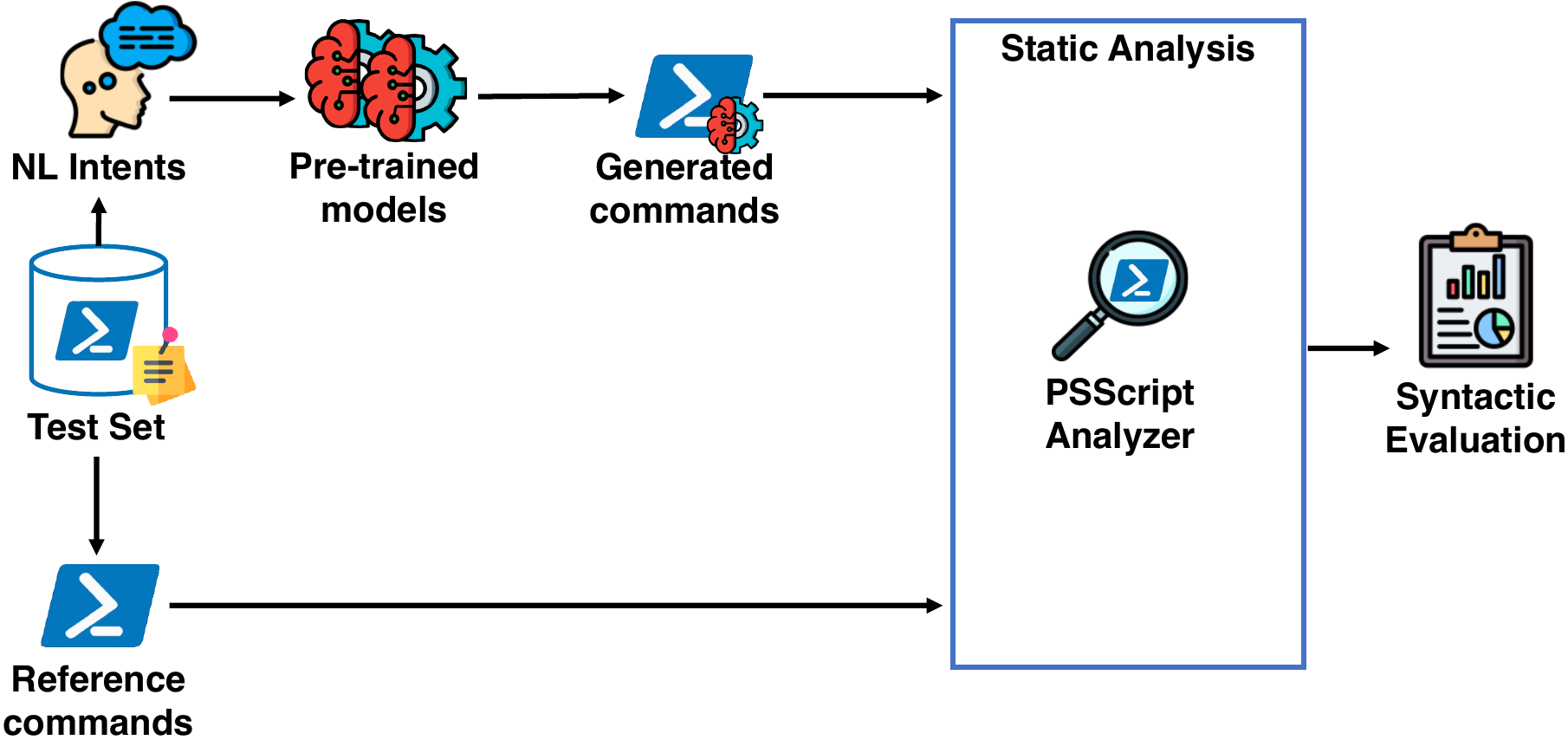}
    \caption{Static analysis workflow.}
    \label{fig:syntax-overview}
\end{figure}

\begin{table}[t]
\centering
\small
\begin{tabular}{
>{\centering\arraybackslash}m{2cm} |
>{\centering\arraybackslash}m{2.5cm} 
>{\centering\arraybackslash}m{2.5cm}} 
\toprule
\textbf{Model} & \textbf{Single Accuracy (\%)} & \textbf{Comparative Accuracy (\%)}\\
\midrule
    CodeT5+ & 91.15 \DrawPercentageBar{0.9115} & 92.04 \DrawPercentageBar{0.9204} \\
    CodeGPT & 98.23 \DrawPercentageBar{0.9823} & 98.23 \DrawPercentageBar{0.9823} \\
    CodeGen & 98.23 \DrawPercentageBar{0.9823} & 98.23 \DrawPercentageBar{0.9823} \\
\bottomrule
\end{tabular}
\caption{Syntactic evaluation for the best models.}
\label{tab:syntax-results}
\end{table}

We developed a syntactic analysis tool to streamline the process of detecting \textit{parse errors}, \textit{errors}, and \textit{warnings} in PowerShell scripts. This tool automatically feeds PSScriptAnalyzer with PowerShell commands generated by the models during the testing phase. By doing so, our tool identifies errors and warnings in the generated code, assessing the overall syntactic quality of the models.

The syntactic analysis process begins with our test set, which consists of NL intents paired with reference PowerShell commands. These NL intents are fed into fine-tuned models to produce the PowerShell code. Both the generated commands and their corresponding references are then subjected to the syntax analyzer.

To assess the syntactic quality of the generated commands, we introduce two distinct metrics: \textit{Single Syntax Accuracy} and \textit{Comparative Syntax Accuracy}. The metrics are defined as follows:

\begin{itemize}
    \item \textbf{Single Syntax Accuracy}: evaluates the percentage of commands without parse errors. This evaluation is independent of the reference commands from the ground truth.
    
    \item \textbf{Comparative Syntax Accuracy}: assesses the syntactic correctness of the generated commands by considering the results alongside the reference commands. When both commands present common parse errors, these are excluded from the counting process. Given that some reference commands include stub templates such as \texttt{<code>} or \texttt{<command>}, the analysis filters out parse errors associated with these templates, specifically the \textit{RedirectionNotSupported} and \textit{MissingFileSpecification} errors.
\end{itemize}

\begin{figure}[t]
    \centering
     \includegraphics[width=1\columnwidth]{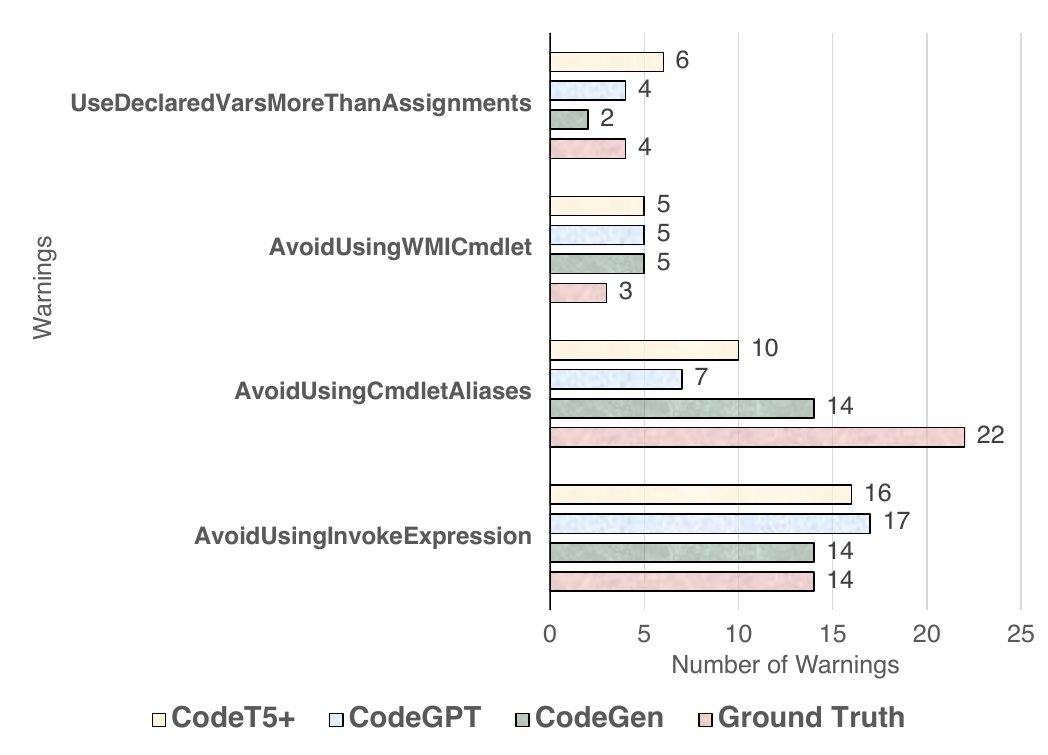}
    \caption{Counts for different warning types in each test set.}
    \label{fig:count-warnings}
\end{figure}

The workflow for the syntactic analysis is depicted in Figure \ref{fig:syntax-overview}. Looking at the results in Table \ref{tab:syntax-results}, it is possible to notice that all the models achieved a score greater than 90\%, assessing their strong capability to generate syntactically correct code. CodeGPT and CodeGen, in general, demonstrate high performance across both syntax metrics. Table \ref{tab:summarysyntax} summarizes the percentages for various severity types in the test set. Given that warning frequencies are consistently above 30\% for all models, including the ground truth, Figure \ref{fig:count-warnings} enumerates the various warning types within each set.

\begin{table}[t]
\renewcommand{\arraystretch}{1.1}
\centering
\makebox[\columnwidth]{% 
\centering
\small
\begin{tabular}{lccc}
\toprule
\textbf{Test Set} & \textbf{ParseError (\%)} & \textbf{Error (\%)} & \textbf{Warning (\%)} \\
\midrule
CodeT5+ & 8.85 \DrawPercentageBar{0.0885} & 1.94 \DrawPercentageBar{0.0194} & 35.92 \DrawPercentageBar{0.3592} \\
\midrule
CodeGPT & 1.77 \DrawPercentageBar{0.0177} & 2.70 \DrawPercentageBar{0.027} & 29.73 \DrawPercentageBar{0.2973} \\
\midrule
CodeGen & 1.77 \DrawPercentageBar{0.0177} & 1.80 \DrawPercentageBar{0.018} & 31.53 \DrawPercentageBar{0.3153} \\
\midrule
Ground Truth & 2.65 \DrawPercentageBar{0.0265} & 0.00 \DrawPercentageBar{0.0} & 39.09 \DrawPercentageBar{0.3909} \\
\bottomrule
\end{tabular}
}
\caption{Summary of ParseError, Error, and Warning percentages for models and ground truth on the test set.}
\label{tab:summarysyntax}
\end{table}

\begin{comment}
\begin{figure}[htbp]
    \centering
     \includegraphics[width=\columnwidth]{img/Warning Frequencies.pdf}
    \caption{Counts for different warning types in each test set.}
    \label{fig:count-warnings}
\end{figure}
\end{comment}

\subsection{Execution Analysis}
\label{subsec:execution}

\begin{figure}[t]
    \centering
    \includegraphics[width=1\columnwidth]{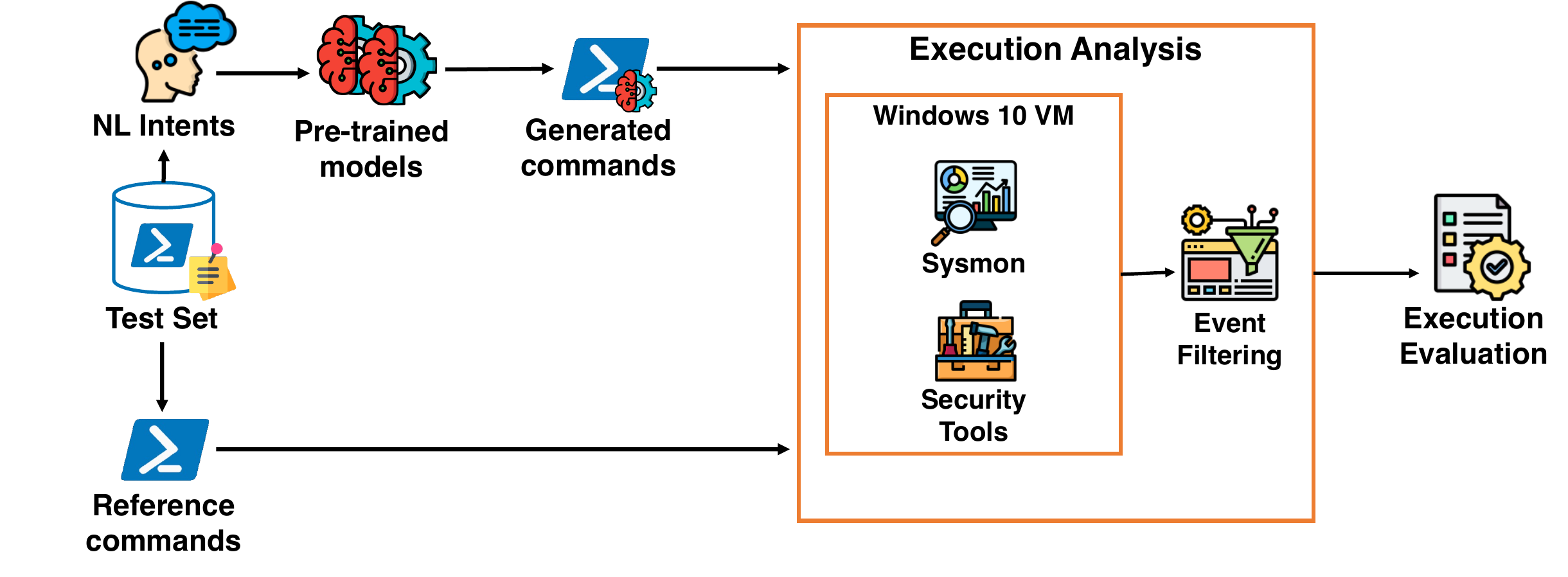}
    \caption{Execution analysis workflow.}
    \label{fig:pwsh-exec}
\end{figure}

The execution analysis aims to evaluate the generated offensive PowerShell code when running in an actual system. This involves assessing the ability of the code to behave as intended in terms of effects caused on the system. Therefore, we run both code from the ground truth and generated code, monitor their behavior at runtime, and compare the behavioral events that occurred during their execution. The entire workflow for the execution analysis is shown in Figure \ref{fig:pwsh-exec}.  

%\roberto{Aggiungere qui esempi di comandi ed eventi}
\begin{figure}[t]
    \centering
    \includegraphics[width=1\columnwidth]{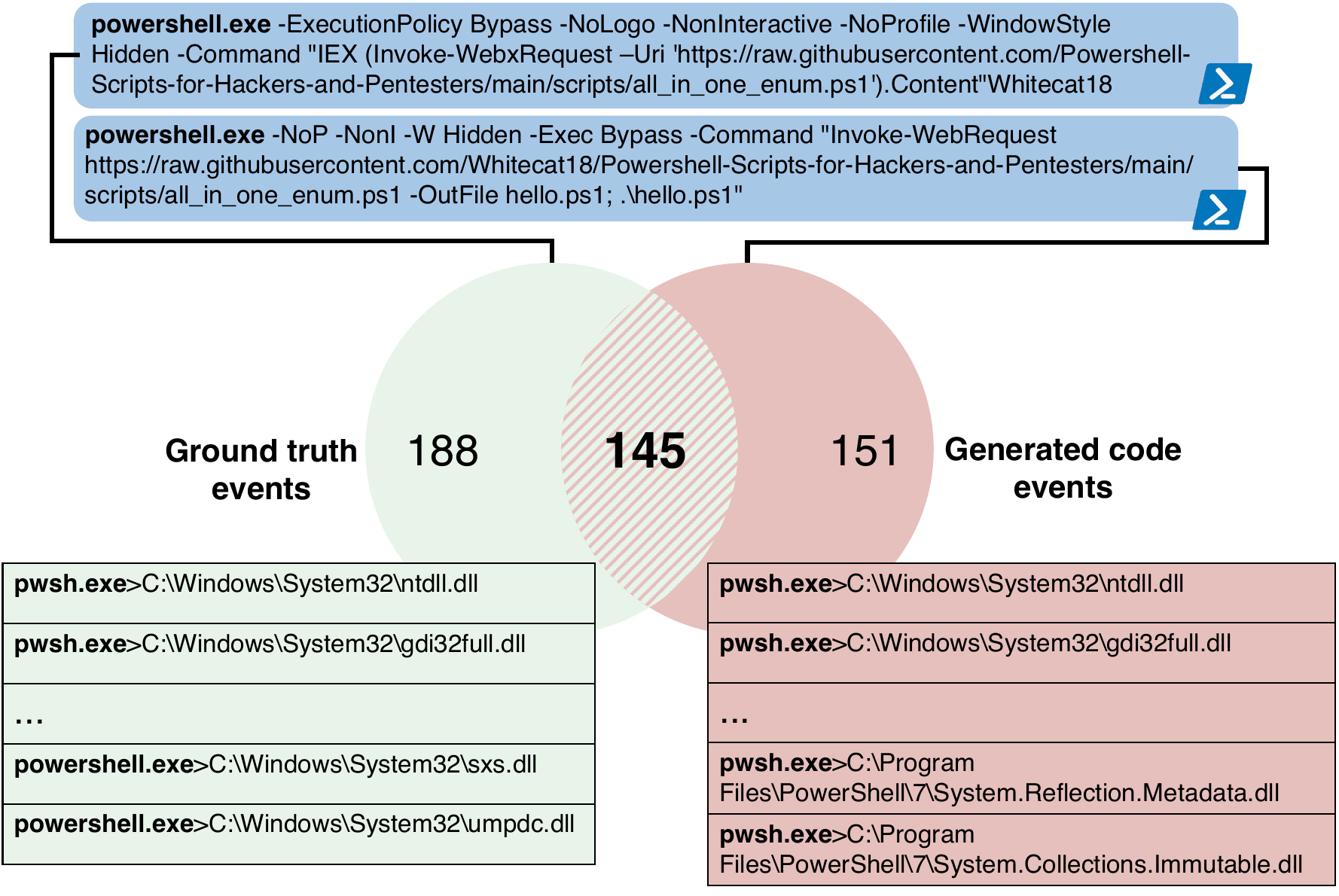}
    \caption{Comparison between events.}
    \label{fig:Events Examples}
\end{figure}

We performed the experiments in a controlled and dedicated testing environment. The controlled environment consists of a virtualized Windows 10 system running in VirtualBox 7. The system is equipped with a set of security-related tools, such as PowerSploit~\cite{powersploit} and Mimikatz~\cite{mimikats}, that are invoked by many samples of offensive code in our dataset. We assume that these tools have been previously infiltrated by the attacker in a previous stage, as typical of advanced malicious campaigns. 
%We focus on evaluating 35 entries from the test set, \textcolor{red}{which have full syntactic correctness of the chosen commands in the ground truth, span different attack scenarios and objectives, and that we could run in our virtualized environment that did not require the setup of external services}. 
To monitor the execution of Powershell code, we integrated Sysmon~\cite{sysmon}, a popular Windows service for gathering system events, including the filesystem, the network, and the Windows Registry. To be able to run the generated code on the system, we assume the scenario in which an attacker already bypassed part of the security mechanisms by deactivating the Microsoft Defender Firewall, Windows Defender, and Microsoft Defender SmartScreen.

%\roberto{dire che PowerSploit e Mimikatz sono tool richiamati da molti programmi nel nostro dataset, e quindi li includiamo per rendere eseguibili gli attacchi. Inoltre, per poter eseguire gli attacchi, abbiamo disattivato alcuni meccanismi di sicurezza come Windows Defender, the firewall, and SmartScreen, assumendo che l'attaccante li abbia evasi o disattivati con altri coman di. Sysmon viene usato ai fini di monitorare gli eventi durante le esecuzioni dei programmi. Sysmon è un tool popolare di analisi dinamica, in grado di raccogliere eventi riguardanti l'accesso alle risorse di sistema tra cui il filesystem, la rete, e il registro di Windows.
%\roberto{Per valutare il comportamento dinamico del codice, confronteremo gli eventi dell'esecuzione del codice generato con gli eventi dell'esecuzione del codice della ground truth, comparando sia il tipo di eventi che le risorse accedute. Per ridurre gli errori nel confronto dovuti a eventi non-deterministici (ad esempio, il codice può ripetere una operazione più volte in caso di delay o errore), il codice è stato eseguito più volte, e sono stati rimossi gli eventi che si verificano sporadicamente. Le esecuzioni sono state effettuate all'interno di una macchina virtuale con Windows 10 su VirtualBox 7. Prima di ogni esecuzione, lo stato pulito della macchina virtuale è stato ripristinato a partire da uno screenshot, per evitare l'interferenza di effetti residui a causa di precedenti programmi malevoli.}
%\luciano{aggiunto - vedere se va bene}

The evaluation involved executing each command from both the generated ones and those from ground truth multiple times as a single-line PowerShell script. This generates a process through the standard Windows \textit{System.Diagnostics.Process}. We filter the events recorded by Sysmon by filtering out records related to previous irrelevant events and selecting records based on the Process ID (PID), focusing on both the parent process responsible for executing the PowerShell command and its child processes. The comparison has been performed comparing the events triggered by the generated command (called \emph{retrieved records}) to those from the execution profile of the ground truth (called \emph{relevant records}). The events that appear both when executing the generated code and the ground truth are \textit{relevant records retrieved}. From these sets of events, we evaluate the \emph{precision}, \emph{recall} and \emph{F1-score} of the generated code, defined as follows:

\[\mbox{precision}= \frac{1}{N} \sum_i^N \frac{\#(\mbox{relevant records retrieved})_i}{\#\mbox{(retrieved records})_i} \]
\[\mbox{recall}= \frac{1}{N} \sum_i^N \frac{\#(\mbox{relevant records retrieved})_i}{\#\mbox{(relevant records})_i} \]
\[ \mbox{F1-Score}= 2\hspace{2pt}\frac{\mbox{precision}*\mbox{recall}}{\mbox{precision}+\mbox{recall}} \]

Figure \ref{fig:Events Examples} illustrates an example of event analysis: given the ground truth and the generated PowerShell command, we execute them and compare the set of events triggered by each command to measure their overlap. To avoid noise in the analysis due to events that only occur sporadically (e.g., because of non-determinism sources in the system), we identify such events by performing multiple repeated runs of the code and discard non-reproducible events from the analysis. After every command execution, the Windows environment is restored to a clean state, by reloading the virtual machine from a snapshot, to avoid interferences caused by the effect of previous commands.

\begin{table}[t]
\renewcommand{\arraystretch}{1.1}
\centering
\makebox[\columnwidth]{% 
\centering
\small
\begin{tabular}{lccc}
\toprule
\textbf{Model} & \textbf{Precision (\%)} & \textbf{Recall (\%)} & \textbf{F1-Score (\%)} \\
\midrule
CodeT5+ &  97.26 \DrawPercentageBar{ 0.9726} &  80.94 \DrawPercentageBar{0.8094} &  88.35 \DrawPercentageBar{0.8835} \\
\midrule
CodeGPT &  91.86 \DrawPercentageBar{0.9186} & 85.23 \DrawPercentageBar{0.8523} &  88.42 \DrawPercentageBar{0.8842} \\
\midrule
CodeGen &  96.94 \DrawPercentageBar{0.9694} & 80.97 \DrawPercentageBar{0.8097} &  88.24 \DrawPercentageBar{0.8824} \\

\bottomrule
\end{tabular}
}
\caption{Execution analysis results.}
\label{tab:execution-analysis-res}
\end{table}

The results shown in Table~\ref{tab:execution-analysis-res} outline how all models share an overall precision higher than 90\% and an overall recall higher than 80\%, likewise, the Execution F1-Score is very similar between the different models and higher than 88\%. 
Thus, although there were differences found in the textual similarity analysis, the generated code closely matches the ground truth in terms of dynamic events.

%\luciano{Aggiunto diagramma di eulero con qualche evento, onestamente toglierei i comandi sotto e le farei sotto forma di tabelle (a parte) essendo che un evento viene identificato da due campi quello di 'Image'(in grassetto in figura) e quello di 'image loaded', anche perché risulta poco leggibile }
\begin{mybox}{\parbox{8cm}{RQ3: How good is the generated code in terms of code quality and dynamic behavior?}}
The syntactic analysis of the generated code showed that the models are indeed capable of generating high-quality PowerShell code. CodeGPT and CodeGen achieve the best results in terms of Single and Comparative Accuracy, along with an amount of Warnings and ParseErrors comparable to the ground truth. The execution analysis revealed that the generated PowerShell code closely replicates the behavior of the ground truth code, generating the same events in the target system. This is indicative of the generated code's capability of performing the malicious actions described in the NL intents. 
\end{mybox}

\subsection{Comparison with Public AI Model}
\label{subsec:comparison}
In this study, we conducted a comprehensive evaluation by comparing the performance of our fine-tuned models, CodeT5+, CodeGPT, and CodeGen, with ChatGPT, the OpenAI LLM service widely used for a variety of tasks, including code generation~\cite{dong2023self}. 
The purpose was to assess the specialized capabilities of our models in generating PowerShell code for offensive security tasks and to benchmark their performance against a publicly available, closed-source model. We leveraged ChatGPT 3.5, which represents the most recent free version at the time of this work. 

To assess the capabilities of the OpenAI model, we first provided a detailed description of the required task, i.e., the generation of PowerShell commands starting from NL descriptions, including an example of input and the desired output. Then, we provided a list of natural language code descriptions and asked ChatGPT to automatically generate the corresponding PowerShell code. Specifically, following works and guidelines on prompt engineering \cite{dong2023self, MicrosoftPrompt}, we leveraged the following prompt: \texttt{I want you to act as a code generator. Given a natural language description of a PowerShell command, generate the corresponding PowerShell code}.

\begin{figure}[t]
    \centering
    \includegraphics[width=1\columnwidth]{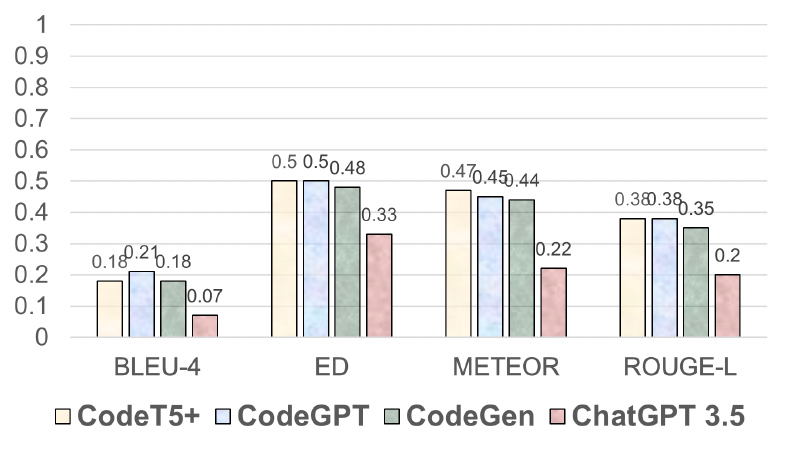}
    \caption{Comparison with ChatGPT on output similarity metrics.}
    \label{fig:cmp-gpt-sim}
\end{figure}

\figurename{}~\ref{fig:cmp-gpt-sim} shows the results of this analysis. The figure shows that our fine-tuned models consistently outperform ChatGPT across multiple evaluation metrics. Specifically, ChatGPT exhibits a BLEU-4 score of 7.45\%, an ED of 33.84\%, a METEOR of 22.14\%, and a ROUGE-L of 20.61\%. In contrast, our fine-tuned models showcase superior overall performance across all output similarity metrics. The tailored training on the specialized fine-tuning dataset, designed specifically for offensive security code generation, results in more accurate code generation, enabling our models to surpass the capabilities of ChatGPT in this particular task.
We also analyzed the syntactical quality of the PowerShell code generated by ChatGPT, obtaining a Syntax Single Accuracy of 95.58\% and a Syntax Comparative Accuracy of 96.46\%. These results underscore the commendable ability of ChatGPT to generate accurate and syntactically correct PowerShell code. 

\begin{figure}[t]
    \centering
    \includegraphics[width=1\columnwidth]{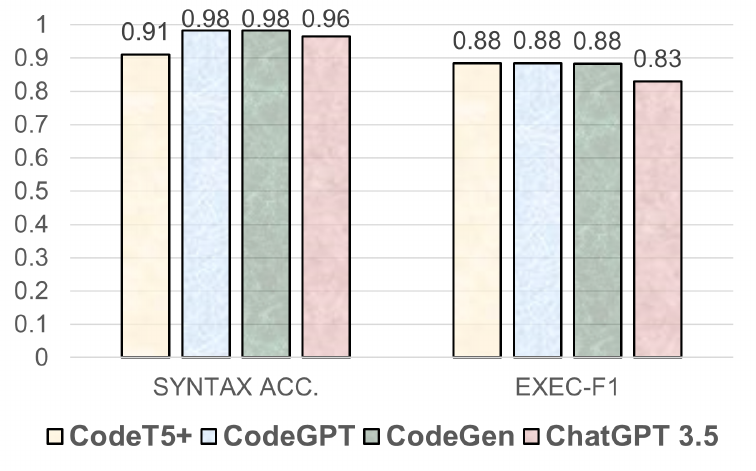}
    \caption{Comparison with ChatGPT on static and execution analysis evaluation metrics.}
    \label{fig:cmp-gpt-exe}
\end{figure}

Finally, we extended the execution analysis to ChatGPT, following the same strategies described in Section \ref{subsec:execution}, obtaining an overall Execution F1-Score of 82.92\%. Despite the strong syntactic performance, ChatGPT remains one step below the fine-tuned models in the qualitative analysis of the generated PowerShell code. The results of this analysis are shown in Figure \ref{fig:cmp-gpt-exe}.

%\luciano{ \ref{fig:cmp-gpt-exe}è nel folder immagini è presente anche quella che va da 0.75 a 100}
\begin{mybox}{\parbox{8cm}{RQ4: How do fine-tuned NMT models, leveraging security-oriented training data, compare to a publicly available, closed-source model?}}
The comparative analysis with ChatGPT, a publicly available general-purpose language model, highlights the specialized strengths of privately fine-tuned models, CodeT5+, CodeGPT, and CodeGen, in offensive PowerShell code generation. The fine-tuned models consistently outperform ChatGPT across BLEU-4, Edit Distance, and METEOR scores. While showing notable performance on syntactic accuracy, ChatGPT achieves poorer results than the fine-tuned models for the execution analysis. This underscores the significance of domain-specific fine-tuning and the benefits of training on security-oriented datasets, providing an advantage in generating offensive PowerShell code compared to a general-purpose language model. The results affirm the effectiveness of tailored training data for achieving superior performance in domain-specific tasks.
\end{mybox}

\section{Threats To Validity}
\label{sec:threats}

\noindent
\textbf{Model selection.} The external validity of the study might be impacted by the choice of NMT models (CodeT5+, CodeGPT, CodeGen). To mitigate this, we carefully selected models with distinct architectures and capabilities, ensuring a representation of current advancements in the field~\cite{lu2021codexglue, wei2023copiloting, tipirneni2022structcoder}. 
This careful selection aims to ensure that our findings reflect broader trends in NMT model performance for code generation tasks.
%Future work will explore a broader spectrum of models for a more comprehensive assessment.

%\vspace{0.1cm}
\noindent
\textbf{Evaluation metrics.} %The use of output similarity metrics might present a threat to construct validity because they may not fully capture the correctness of the generated Powershell commands. However, we remark that this is the common practice for the evaluation of NMT models since semantics evaluation of the code requires human effort, hence posing practical challenges for large-scale assessments~\cite{LIGUORI2023120073}.
The reliance on output similarity metrics, although representing the most common solution in the field, poses a potential threat to construct validity, as these metrics may not fully encapsulate the correctness and functional adequacy of the generated PowerShell commands. To address this issue, our evaluation strategy encompasses a comprehensive suite of metrics, including similarity, syntactic, and execution metrics, each offering unique insights into the models' performance. By considering multiple variants of these metrics and aligning with common practices in code generation evaluation, we aim to provide a well-rounded assessment. No single metric is perfect, but analyzing them collectively allows for a more comprehensive evaluation of the code.

%\vspace{0.1cm}
\noindent
\textbf{Fine-tuning data.} %Our dataset construction involved meticulous curation from diverse sources, including online repositories and guides, Atomic Red Team, Stockpile, Empire, and ChatGPT API interactions. However, the specific characteristics of the selected sources might limit the generalizability of the models' performance to other offensive security contexts. To address this, we aimed for a comprehensive representation of PowerShell commands used in cybersecurity. Additionally, to enhance the dataset's relevance, NL descriptions were crafted following styles found in PowerShell books~\cite{TutorialsPointPowerShell, lee2011windows, holmes2012windows}, and descriptions were left unchanged when already embedded within the code. This approach aligns with real-world scenarios where NL descriptions are used to articulate the purpose of PowerShell commands.
The construction of our dataset, meticulously curated from several sources such as online repositories, Atomic Red Team, Stockpile, and Empire, introduces potential limitations regarding the generalizability of our models' performance across different offensive security contexts. 
To minimize the impact of these limitations, we sourced data from diverse origins and conducted manual verification of each sample in the labeled dataset, ensuring the completeness and coherence of descriptions with the intended programs. The diversity in data sources and the thorough verification process aim to diminish the influence of any singular source's peculiarities and errors in programs or descriptions, thereby enhancing the dataset's applicability and reliability for training and evaluating AI models in generating offensive PowerShell code. 
Furthermore, our approach to crafting NL descriptions, inspired by established styles found in PowerShell literature, mirrors real-world scenarios where such descriptions play a critical role in describing PowerShell commands. 
Finally, regarding the size of our dataset, we notice that it is in line with other state-of-the-art corpora used to fine-tune models, which are in the order of one thousand samples~\cite{zhou2023lima}.
\section{Ethical Considerations}
\label{sec:ethics}

Recognizing that attackers use attacks as a weapon, it is important to specify that the goal of the proof-of-concept (POC) is not to cause harm but to surface security weaknesses within the software. Identifying security issues allows companies to patch vulnerabilities and protect themselves against attacks. 

\emph{Offensive security} is a sub-field of security research that tests security measures from an adversary or competitor’s perspective, employing ethical hackers to probe a system for vulnerabilities~\cite{bratus2013offensive,oakley2019state}. 
Our work aims to automate attack generation to explore critical vulnerabilities before they are exploited by attackers~\cite{avgerinos2011aeg}. 
Indeed, our work simplifies the process of coding the attacks to surface security weaknesses within the software and can provide valuable information about the technical skills, degree of experience, and intent of the attackers. With this information, it is possible to implement measures to detect and prevent attacks~\cite{arce2004shellcode}.

\section{Conclusion}
\label{sec:conclusion}

\normalsize In this paper, we assessed the feasibility of using NMT models to generate PowerShell code for security contexts. We aimed to demonstrate that AI-based code generators are indeed fit to generate PowerShell code, specifically, offensive PowerShell, which spans several applications in the cybersecurity domain. The evaluation of CodeT5+, CodeGPT, and CodeGen demonstrated that these models achieve significant performance on the code generation task, both with and without pre-training. Moreover, the study showed that domain-specific fine-tuning allows our models to outperform state-of-the-art privately fine-tuned models, i.e., ChatGPT. We also introduced two novel datasets for PowerShell code generation to use for pre-training and fine-tuning AI-code generators.

Future work includes further analysis of the generated code, such as sandbox execution of the offensive scripts, to understand whether the code can evade detection measures, along with more NMT models spanning several architectures and capabilities.

%\pietro{manca una frase sul dataset}

%\pietro{metti un piccolo paragrafo di future work, dove diciamo che faremo un'analisi anche manuale, tipo esecuzione attacchi, e che includeremo piu modelli (vedi threats)}
\section*{Acknowledgments}
%\footnotesize 
This work has been partially supported by MUR PRIN 2022, project FLEGREA, CUP E53D23007950001 (\url{https://flegrea.github.io}) and by an Industrial Ph.D. grant (PNRR - DM 117/2023) from MUR and DigitalPlatforms S.p.A, CUP E66E23000580003. %\normalsize

\bibliographystyle{IEEEtran}
\bibliography{biblio}

%%%%%%%%%%%%%%%%%%%%%%%%%%%%%%%%%%%%%%%%%%%%%%%%%%%%%%%%%%%%%%%%%%%%%%%%%%%%%%%%
\end{document}